\documentclass[journal]{IEEEtran}

\usepackage{cite}
\usepackage{amsmath}
\usepackage{array}
\ifCLASSOPTIONcompsoc
  \usepackage[caption=false,font=normalsize,labelfont=sf,textfont=sf]{subfig}
\else
  \usepackage[caption=false,font=footnotesize]{subfig}
\fi

\usepackage{fixltx2e}

\usepackage{url}

\usepackage{color}  
\definecolor{orange}{rgb}{1,0.5,0}
\usepackage{booktabs} \newcommand{\ra}[1]{\renewcommand{\arraystretch}{#1}}

\usepackage{epsf}
\usepackage{epsfig} 
\usepackage[newfloat]{minted}

\SetupFloatingEnvironment{listing}{name=Prog.} 
\setminted[c]{ 
fontsize=\footnotesize, 
escapeinside=||, 
mathescape=true, 
numbersep=5pt, 
gobble=1, 
linenos=true, 
framesep=3mm}

\usepackage{listings}

\lstdefinelanguage{customc}{
  keywords={for},
  keywordstyle=\color{blue}\bfseries,
  keywordstyle=[2]\color{red}\bfseries,
  keywordstyle=[3]\bfseries,
  keywordstyle=[4]\itshape,
  belowcaptionskip=1\baselineskip,
  identifierstyle=\color{black},
  breaklines=true,
  numbersep=8pt,
  xleftmargin=.2in,
  showstringspaces=false,
  basicstyle=\footnotesize\ttfamily,
  commentstyle=\itshape\color{purple},
  stringstyle=\color{orange},
}
\usepackage{mathtools} 
\usepackage{amssymb}
\usepackage{bbding}
\usepackage{bm}

\usepackage{algorithm}     
\usepackage{algorithmicx}  
\usepackage{algpseudocode} 
\usepackage{varwidth}      

\newcommand{\A}{\texttt A}
\newcommand{\C}{\texttt C}
\newcommand{\G}{\texttt G}
\newcommand{\T}{\texttt T}

\newcolumntype{L}{>{\raggedright\arraybackslash}X}
\newcolumntype{s}{>{\hsize=.9\hsize}X}

\begin{document}
\title{A High-Efficiency SoC for Next-Generation Mobile DNA Sequencing}

\author{Abel Beyene,~\IEEEmembership{Graduate Student Member,~IEEE,}
        Zhongpan Wu,
        Yunus Dawji,
        Karim Hammad,~\IEEEmembership{Member,~IEEE,}\\
        Ebrahim Ghafar-Zadeh,~~\IEEEmembership{Senior Member,~IEEE,}
        and~Sebastian Magierowski
\thanks{
The authors are with the Department
of Electrical Engineering and Computer Science, York University, Toronto,
ON M3J 1P3, Canada (e-mail: magiero@eecs.yorku.ca).

Karim Hammad is with the Arab Academy for Science, Technology and Maritime Transport, Cairo, Egypt (e-mail: khammad@aast.edu).
}}

\maketitle

\begin{abstract}
Hand-sized Deoxyribonucleic acid (DNA) sequencing machines are of growing importance in several life sciences fields as their small footprints enable a broader range of use cases than their larger, stationary counterparts. However, as currently designed, they lack sufficient embedded computing to process the large volume of measurements generated by their internal sensory system. As a consequence, they rely on external devices for additional processing capability. This dependence on external processing places a significant communication burden on the sequencer's embedded electronics. Moreover, it also prevents a truly mobile solution for sequencing in real-time. Anticipating next-generation machines that include suitably advanced processing, we present a System-on-Chip (SoC) fabricated in 22-nm complementary metal-oxide semiconductor (CMOS).  Our design, based on a general-purpose reduced instruction set computing (RISC-V) core, also includes accelerators for DNA detection that allow our system to demonstrate a 13X performance improvement over commercial embedded multicore processors combined with a near 3000X boost in energy efficiency.
\end{abstract}


 \ifCLASSOPTIONpeerreview
 \begin{center} \bfseries EDICS Category: 3-BBND \end{center}
 \fi

\IEEEpeerreviewmaketitle

\section{Introduction}\label{s:introduction}
\subsection{Background and Motivation}
\IEEEPARstart{D}{NA} sequencing has undergone a transformative evolution since its inception. Early approaches, such as Sanger sequencing, relied on chain-termination methods and were characterized by high accuracy but limited throughput and scalability \cite{sanger1977dna}. The advent of next-generation sequencing (NGS) technologies, including Illumina and pyrosequencing platforms, revolutionized the field by enabling massively parallel sequencing and significantly reducing sequencing time and costs \cite{8094922,9063002}. More recently, third-generation sequencing (TGS) technologies, particularly nanopore-based approaches, have enabled long-read Deoxyribonucleic acid (DNA) sequencing, real-time data acquisition, and portable operation \cite{quick2016real}. Mobile sequencing, epitomized by handheld devices like the Oxford Nanopore Technologies MinION, is increasingly adopted for in-field applications, including real-time pathogen detection, epidemiological surveillance, and environmental monitoring \cite{Loit19Agri}. Furthermore, mobile sequencing plays an expanding role in personalized health and precision medicine by enabling point-of-care diagnostics, pharmacogenomics, and rapid identification of actionable genetic variants for individualized treatment strategies \cite{dyshlovoy2024applications}. Despite their portability and versatility, these devices face computational challenges due to limited on-board processing capabilities, necessitating embedded bioinformatics solutions such as the System-on-Chip (SoC) design proposed in this paper to enable efficient, real-time DNA sequence detection in mobile contexts.

DNA sequencers have experienced a dramatic reduction in size.  Today, sequencing machines about the volume of a smartphone are available for less than \$1,000~\cite{Oliva20}.  These miniature devices are also fast and, at maximum sustained throughput, have the ability to measure the equivalent of a human genome in about four hours within a 5-W power budget~\cite{Clarke19}.  Importantly, these machines produce their results in real-time and respond quickly to new sample inputs.  Being relatively new, miniature sequencing technology holds significant potential for achieving further footprint reductions and for reaching further DNA measurement throughput boosts.  

\begin{figure}[h]
\centerline{\includegraphics[width=3.4in]{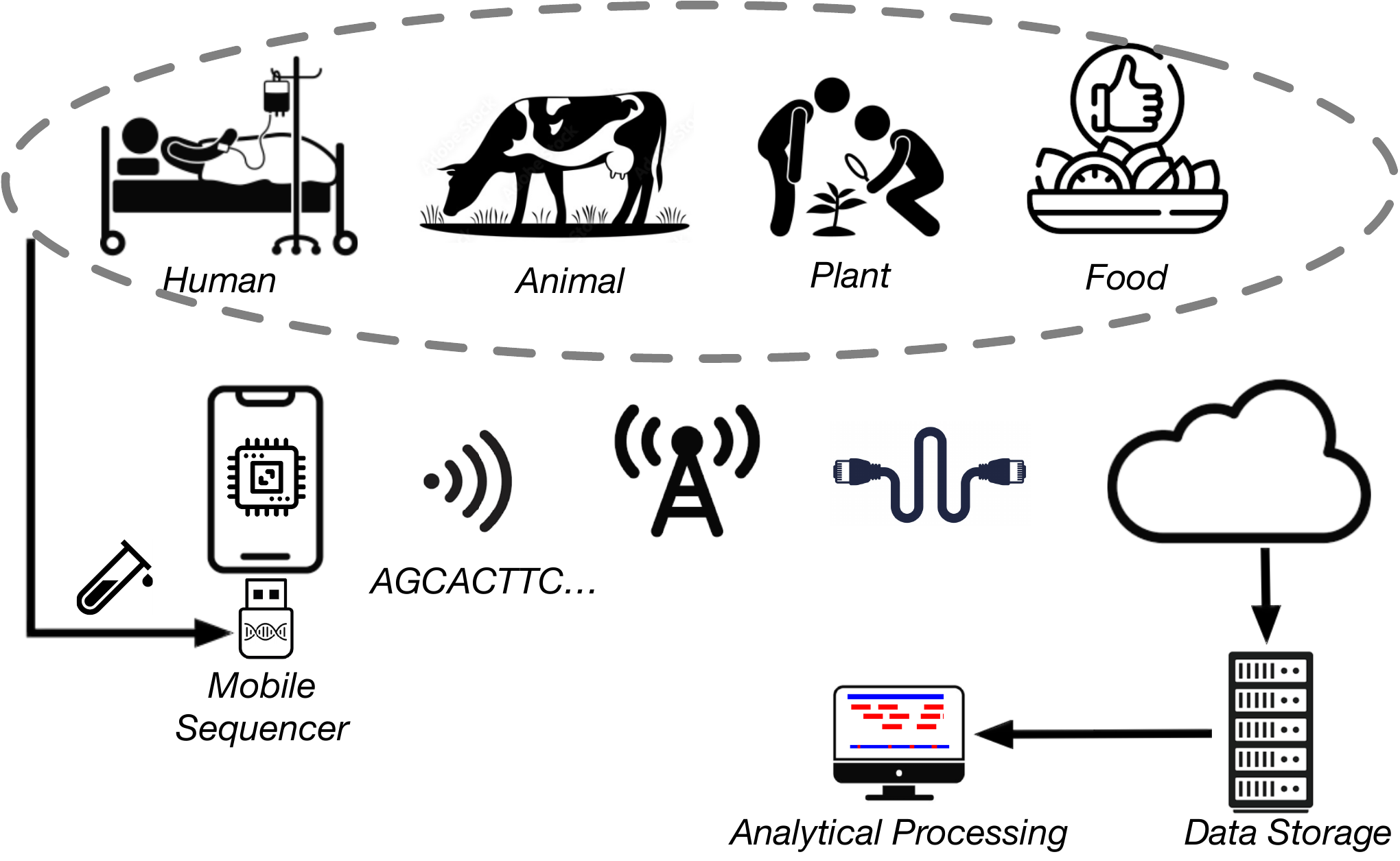}}
  \caption{An example of miniature DNA sequencer within an IoT genomic sensing system.}
\label{f:MobSeq}
\end{figure}

With such characteristics, the practical applications addressable by miniature sequencers are likely to grow beyond their current adoption in the genomic research space and towards personalized medicine, public health monitoring, agricultural testing, industrial processing, and even next generation information technology~\cite{Wang21,Gorzynski22,Loit19Agri,Chen20,Liu21}.  
These characteristics also present intriguing opportunities for Internet of Things (IoT) applications.  Such a scenario, inspired by emerging online water monitoring solutions~\cite{Manjakkal2021iotj}, is imagined in Fig.~\ref{f:MobSeq}.  Therein, small mobile sequencers allow sophisticated genomic data to be gathered from numerous environmental sources and wirelessly forwarded to remote processing centres for deeper interpretation and analysis.

Innovations in sensors~\cite{Verren20} and read-out electronics~\cite{Dawji21} serve as the primary technological underpinnings behind the success of miniature DNA sequencers to-date~\cite{Asgari25}.  At present however, these machines do not include any significant amount of computing within their small chassis.  Rather, the raw DNA measurements they gather (which come in the form of small electrochemical current signals) are forwarded to external computing devices like desktops or laptops.  These external computers then carry out follow-up data analysis or simply store the measurements before sending them to downstream cloud facilities~\cite{Xu20}. 

\subsection{Challenges and Design Goals}

Alas, currently the recommended sequencer system configuration relies on a 500-MB/s USB link~\cite{Kchouk2015} from the sequencer to external computing or data storage.  For proper operation, this link must be capable of maintaining uninterrupted communication.  As a result, present communication needs greatly compromise the suitability of miniature DNA sequencers for possible IoT application scenarios as outlined in Fig.~\ref{f:MobSeq}.

However, if sufficiently adequate computing is embedded directly within miniature sequencers, this additional processing resource can be used to complete select analyses directly within the sequencer itself.  As a result, the problem of communicating unfiltered broadband measurement data to downstream computing may be greatly alleviated.  

For instance, if the sequencer's physical measurements (e.g., electrochemical current signals) are computationally transformed to predictions of the molecule sequences (i.e., known as \emph{basecalling}) that generated them (i.e., FastA data format) -- in the case of DNA this would require just two bits for each possible monomer adenine (\A), cytosine (\C), guanine (\G), thymine (\T) -- a ${\sim}$100$\times$ bandwidth compression can be realized compared to the Fast5 format (i.e., used with external computing or storage) \cite{Sikolenko2021}.  If these translations are also subject to further analysis (e.g., communicating only relevant monomer sequences for some detection problem), a further 100$\times$ bandwidth reduction may be realized and even the lifetime of the sequencer's sensory apparatus may be extended~\cite{Loose16}.  In conjunction, having more control over data interpretation within the sequencing device itself (rather than uploading to centralized facilities) offers a means for increased data privacy and security \cite{Le2025}.

Realizing such benefits from additional embedded computing can help propel emerging miniature DNA sequencing technology towards a much broader set of IoT use cases but only if embedded with sufficiently sophisticated processing \cite{Zheng2023}.  Further, this processing must be achieved without compromising their chief advantage: amenability to mobile applications.  As a result, embedded computers for next-generation IoT DNA sequencers must be small and energy efficient.

To-date, no custom computing designs for miniature sequencing devices have been seriously discussed or any potential candidates experimentally demonstrated in the open literature. Previous works have focused on alternative algorithms~\cite{Sarkar2020,Sharei2024}, simulation-based methodologies~\cite{Mao2022} or the use of commodity hardware~\cite{Wu20,Hammad21,Huang2022,Xu2023}. To underscore the novelty and breadth of our proposed system, Table~\ref{t:Feature_comparison} presents a comparative overview of these solutions and their supported functionalities, including our earlier works. In contrast, this paper presents the first such exploration in this specialized domain by detailing the design, realization, and physical measurement of a complementary metal-oxide semiconductor (CMOS) system-on-chip (SoC) targeting mobile, embedded, bioinformatics applications.  Our proposed system balances flexibility and efficiency by combining heterogeneous processing with bioinformatics-specialized acceleration as well as high-speed interfaces to enable real-time processing. Moreover, it is motivated by a hardware/software co-design approach and evaluates alternative workload partitions directly in silicon.

\begin{table*}
\centering
\caption{Feature Comparison with Existing Hardware Architectures for DNA Analysis.}
\begin{tabular}{lccccccc}
\hline
\textbf{Work} & \textbf{Basecalling} & \textbf{Viterbi-Based} & \textbf{\begin{tabular}[c]{@{}c@{}}Multi-accelerator\\ Architecture\end{tabular}} & \textbf{\begin{tabular}[c]{@{}c@{}}ASIC\\ Implementation\end{tabular}} & \textbf{\begin{tabular}[c]{@{}c@{}}Real-Time\\ Capable\end{tabular}} & \textbf{\begin{tabular}[c]{@{}c@{}}SoC\\ Integration\end{tabular}} & \textbf{Ref.} \\ \hline

Sarkar et al. (IEEE TVLSI, 2020) & \XSolidBrush & \XSolidBrush & \XSolidBrush & \XSolidBrush & \checkmark & \XSolidBrush & \cite{Sarkar2020} \\

Wu et al. (IEEE TBioCAS, 2020) & \checkmark & \checkmark & \XSolidBrush & \XSolidBrush & \checkmark & \XSolidBrush & \cite{Wu20} \\

Hammad et al. (IEEE TVLSI, 2021) & \checkmark & \checkmark & \XSolidBrush & \XSolidBrush & \checkmark & \XSolidBrush & \cite{Hammad21} \\

Mao et al. (IEEE MICRO, 2022) & \checkmark & \XSolidBrush & \XSolidBrush & \XSolidBrush & \checkmark & \XSolidBrush & \cite{Mao2022} \\ 

Huang et al. (IEEE TCBB, 2022) & \checkmark & \XSolidBrush & \XSolidBrush & \XSolidBrush & \XSolidBrush & \XSolidBrush & \cite{Huang2022} \\

Xu et al. (IEEE BIBM, 2023) & \checkmark & \XSolidBrush & \XSolidBrush & \XSolidBrush & \checkmark & \XSolidBrush & \cite{Xu2023} \\

Sharei et al. (IEEE TBioCAS, 2024) & \XSolidBrush & \XSolidBrush & \XSolidBrush & \XSolidBrush & \checkmark & \XSolidBrush & \cite{Sharei2024} \\

Dawji et al. (IEEE ESL, 2024) & \checkmark & \checkmark & \XSolidBrush & \checkmark & \checkmark & \checkmark & \cite{Dawji2024} \\

Proposed SoC & \checkmark & \checkmark & \checkmark & \checkmark & \checkmark & \checkmark & - 
\\  \hline
\end{tabular}
\label{t:Feature_comparison}
\vspace{-0.2cm}
\end{table*}

\subsection{Scope and Contributions}
While our earlier work \cite{Dawji2024} (i.e., added to Table~\ref{t:Feature_comparison}) presented a preliminary SoC prototype for DNA sequence detection, this paper builds upon that foundation with a substantially more advanced architecture with more rigorous experimental evaluation. These contributions mark a significant advancement beyond the initial concept outlined in \cite{Dawji2024}. We summarize the key contributions of this paper as follows:
\begin{itemize}
    \item While our prior work focused on a single-accelerator SoC design for Hidden Markov Model (HMM) trellis construction, this work proposes a more sophisticated multi-accelerator architecture. 
    
    \item We introduce AccelB, a novel accelerator capable of executing both the HMM trellis construction and the memory-intensive traceback algorithm, thereby enhancing the SoC’s performance and energy efficiency.

    \item We use datasets based on predictive nanopore k-mer models to evaluate the proposed SoC’s accuracy across various signal-to-noise ratio (SNR) levels and event chunk sizes.

    \item This work expands the performance benchmarking to include broader comparisons against state-of-the-art designs, including a SIMD-enabled multicore Cortex-A53, a Tensilica Xtensa LX6 reduced instruction set computing (RISC-V) processor, and a desktop-accelerated sequencing platform.
\end{itemize}

Beyond its contributions to the field of bioinformatics, the proposed embedded SoC holds significant potential for applications in the broader consumer electronics (CE) landscape. As mobile health (mHealth) technologies and wearable devices \cite{10554665,le2023securing} continue to evolve, there is an increasing demand for compact, energy-efficient, and high-performance solutions for real-time biomedical data processing \cite{10587030}. Our custom SoC, designed for DNA Viterbi-based sequence detection, could serve as a key enabler in emerging portable health-monitoring and point-of-care diagnostic devices \cite{8502707,sheka2021oxford}. These consumer-facing technologies would benefit from the low-power, high-throughput operation of our SoC, supporting the seamless integration of bioinformatics functions into future CE products.

The organization of our paper is as follows: In Section~\ref{s:sequencing} we describe the IoT system we envisioned, and hence, the hardware and software computing requirements we targeted for our SoC.  Sections~\ref{s:design} and~\ref{s:accelerator} contain detailed descriptions of the proposed SoC design.  Measured performance results are discussed in Section~\ref{s:measurement} followed by a summary and conclusions in Section~\ref{s:summary}.

\section{Embedded Computing for Mobile Sequencing}\label{s:sequencing} 

\subsection{Miniature DNA Sequencing System Description}\label{ss:MiniSeqDescription}

\begin{figure}
\centerline{\includegraphics[width=3.4in]{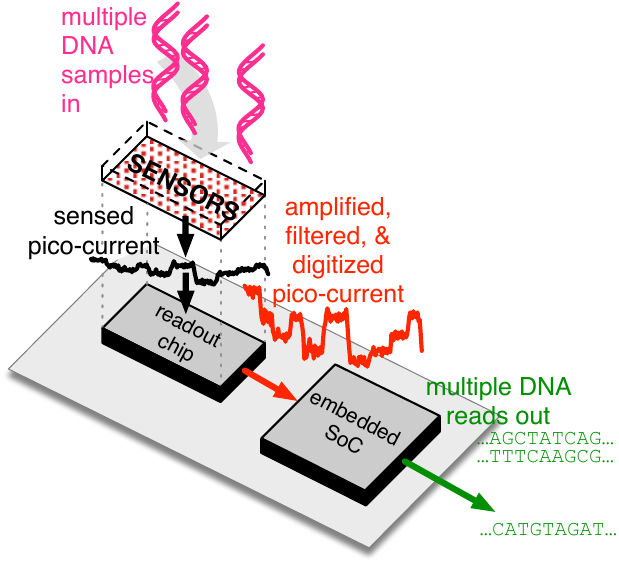}}
  \caption{Component overview of a possible future miniature DNA sequencing system within which more substantial SoC computing is embedded.}
\label{f:app_overview}
\end{figure}
A SoC is a highly integrated computing architecture that consolidates multiple functional units—including processing cores, memory hierarchies, I/O interfaces, and domain-specific accelerators—within a single silicon die to optimize performance, power efficiency, and form factor. SoCs are increasingly employed in lab-on-chip systems \cite{arshavsky2020lab}, where they enable real-time processing and analysis of complex biomedical data, such as DNA sequence detection \cite{Alexandrou21}, within a compact form factor. These benefits are particularly critical for next-generation mobile DNA sequencers, where power efficiency and computational capacity are essential for handling the large data streams produced by nanopore-based sensors \cite{chen2023portable}. The proposed SoC in this work leverages a heterogeneous architecture, including RISC-V cores and dedicated accelerators, to achieve an optimal trade-off between flexibility, computational performance, and energy efficiency, thereby addressing key challenges in mobile sequencing applications.

Fig.~\ref{f:app_overview} sketches the manner in which the SoC described in this work may be integrated within future miniature DNA sequencers intended for service in IoT contexts.  Although a system as imagined in Fig.~\ref{f:app_overview} does not yet exist, it is inspired by the way in which existing miniature DNA sequencers based on nanopore sensors are configured~\cite{Clarke19}.

In existing designs, a parallel array of 1000s of DNA sensors is positioned atop a mixed-signal {\it readout chip}~\cite{Magiero16}.  The sensor array's small ${\sim}1$~cm$^2$ footprint accommodates a compact interface with the readout chip~\cite{Canas14OntPat}.  Each sensor processes only one DNA at-a-time and, in the process, generates a picoampere-scale ionic current (sensed pico-current shown in black in Fig.~\ref{f:app_overview}). The modulations of this minute current are ultimately indicative of the sensed DNA's monomer structure.  Given the parallel nature of the systems, 100s of such pico-current signals can be generated simultaneously and each would be conditioned (amplified, filtered, and digitized) by a separate CMOS channel in the readout chip (not shown)~\cite{Fish21OntPat, Dawji21}. A single conditioned pico-current signal is shown in red in Fig.~\ref{f:app_overview}.

In present-day miniature DNA sequencers (unlike the system sketched in Fig.~\ref{f:app_overview}), the conditioned pico-current signal is serialized within the readout chip and then forwarded (via glue logic) to a neighbouring USB chip~\cite{branton2019nanopore}.  Consequently, the USB chip assembles the data into packets and transmits them out of the miniature sequencer via a wired connection.  As already noted in \S~\ref{s:introduction}, downstream devices then may either process or store the gathered data in hard drives or they may possibly forward it to the cloud for ensuing bioinformatic analysis.

To partially offload dependence on large external computing resources, we propose a system where data from the readout chip is directly processed by an embedded SoC, as illustrated in Fig.~\ref{f:app_overview}.  This embedded computer may then pass its results onto another communications chip (e.g., the aforementioned USB chip), or possibly employ communications blocks of its own.  As noted earlier, by computationally distilling the sequencer's data early in the processing pipeline we greatly reduce the amount of information that needs to be exchanged with downstream remote computing resources.  Embedded computing also opens the door for the device to quickly make higher-level decisions of its own (e.g., tracking high-level patterns in the distilled data).  We now turn to a discussion of the computations that may reasonably be executed on embedded SoCs in future miniature sequencing systems.

Despite their diverse sensing modalities, all DNA sequencers, from large to small, are alike in so far as they produce noisy electronic time-series in response to the DNA samples fed through them.  Due to a number of physical challenges, these measurements are only done on fragmented DNA library samples randomly drawn from locations in longer genomic DNA (gDNA).  As a result, a plethora of algorithms forming a {\it sequencing pipeline} are often needed to operate on the measured data.  In tandem, as part of a computational pipeline, these programs progressively reconstruct relevant parts of animal genomes from which users extract desired insights.  Common examples of computational pipeline steps include mapping, consensus, and assembly were studied in~\cite{Heng18minimap2,Medaka23,Flye22}.  Downstream genomic interpretation also includes steps such as variant calling, annotation, classification~\cite{DeepVariant18,Ensembl22,Centrifuge16}, etc. 

By its nature, a heterogeneous processing system can, to varying degrees of efficiency, be applied to all of these challenges. This motivated our decision to include general-purpose microprocessors in our proposed embedded computing solution. The flexibility afforded by microprocessors, on the other hand, comes with power consumption costs that ultimately compromise the need for energy efficiency in IoT devices.  This can be addressed with specialized hardware accelerators upon which the microprocessor can depend on for specific tasks.

Of the many options to target for hardware acceleration, we have chosen to focus on those tasks that occur at the beginning of the computational sequencing pipeline.  These initial functions are not only computationally intensive, they are also foundational in that many following sequencing algorithms rely on their results~\cite{Lanius2024}~\cite{Sarkar2020}. These early tasks also distill the large amount of data gathered from the DNA into more concise approximations of the measured DNA.  Hence, they afford a reduction in required bandwidth for communication of results to remote processors within which the remainder of a computational pipeline may be completed.

\subsection{Sequence Detection: Trellis Construction}\label{ss:Trellis}

Specifically, our SoC is focused on the implementation of a dynamic programming engine for sequence detection (otherwise known in the literature as \emph{basecalling}).  This refers to the process of converting physical DNA representations (i.e., information from the pico-current time-series discussed in \S~\ref{ss:MiniSeqDescription}) to their text equivalent reads (i.e., the monomer sequence draw from the alphabet $\{\A,~\C,~\G,~\T\}$).

Detection is complicated by the distorted nature of the sensing process, a primary problem being that there is no one-to-one mapping between individual time-series signals (events) and individual monomers, \A, \C, \G, or \T.  Rather, individual time-series events are indicative of some sub-sequence of $k{>}1$ monomers, a $k$-mer.  In essence, $k$ reflects the limited resolution of the sequencer's DNA sensor from which the time-series events are derived.  As a result of this coarse resolution, a sequence detection rather than symbol detection method is needed.  In the DNA processing context, a sequence detector tracks the probable relation between individual events and the space of possible $k$-mers from which these time-series signals could have originated.  These relation probabilities are then used by the detector to derive an estimate of the individual monomer sequence corresponding to the measured time-series sequence.   The result of this process is a monomer text sequence representation of the measurement (a DNA read) which may then be further processed by downstream bioinformatics algorithms (e.g., to carry out alignment).  

Our approach to this sequence detection task employs a Viterbi decoding method common to bioinformatics sequence analysis, a key part of which is expressed with the pseudocode in Fig.~\ref{f:Viterbi}.

\begin{figure}
\begin{lstlisting}
for m = 0 to M-1 {  // *(1) event loop start
  for n = 0 to N-1 { // *(2) state loop start
    *transidx = GatherTrans(n)
    for t = 0 to T-1 { // *(3) trans loop start
      idx = transidx[t]
      trans[t] = @{$\color{red}{\alpha_{m{-}1}}$@[idx] + tprob[t]
    } // end trans loop (3)*
    minidxT = FindMin(trans, T)
    @{$\color{red}{\beta}$}@[n][m-1] = minidxT
    @{\color{red}{$\alpha'_m$}}@[n] = Post(@{\color{red}{$x$}}@[m],trans[minidxT],@{$\mu$}@[n],@{$\sigma$}@[n])
  } // end state loop (2)*
  minidxN = FindMin(@{\color{red}{$\alpha'_m$}}@,N)
  minprob = prob[minidxN]
  for n = 0 to N-1 { // *(4) norm loop start
    @{\color{red}{$\alpha_m$}}@[n] = @{\color{red}{$\alpha'_m$}}@[n]-minprob
  } // end norm loop (4)*
} // end event loop (1)*
\end{lstlisting}
  \caption{Viterbi trellis construction algorithm.}
\label{f:Viterbi}
\end{figure}

The code shown in Fig.~\ref{f:Viterbi} concerns the aforementioned event-to-$k$-mer probability computations.  It does so via a trellis construction phase of the Viterbi sequence detection algorithm.  More specifically, this code's main purpose is to convert a time-series consisting of $M$ input events 
\begin{equation}
\bm{x}{=}\{x\texttt{[m]}\}_{0}^{M{-}1}
\end{equation}
into a $N{\times}M$ matrix (i.e., the trellis) of integer {\it trellis pointers} $\beta\texttt{[n][m]}$.  As explained below, the pointer values denote probable steps through the trellis ``states'' and, in the maximum likelihood sense, can collectively be used to identify the optimal path (i.e., detected sequence) through the trellis.  The $N$ term denotes the number of possible $k$-mers that can be associated with any one event.  In general, since DNA consists of four monomers (\A, \C, \G, \T), the number of possible $k$-mers for a sequencing with sensing resolution of $k$ is $N{=}4^k$.  As shown in Fig.~\ref{f:Viterbi}, for-loop iterations over $M$ and $N$ comprise the two outermost loops of the trellis construction algorithm as it builds the trellis one pointer calculation at-a-time.  An ensuing traceback algorithm (discussed below in \S~\ref{ss:Traceback}) traverses this $N{\times}M$ trellis via the pointers $\beta\texttt{[n][m]}$ to produce the final sequence of monomers that constitute a DNA read.

As noted above, the trellis pointers collectively define a set of optimal paths through the trellis spanning contiguous routes from {\tt m} indexes $M{-}1$ to 0.  More formally, these paths traverse the trellis's $k$-mer {\it state vectors} $\Psi_0,\ldots,\Psi_{M{-}1}$ where $\Psi_m = (\psi_m^0,\ldots,\psi_m^{N{-}1})$ and where each $\psi_m^n$ represents a unique $k$-mer {\it state}\footnote{For example, for $k{=}3$, if the states denote 3-mers in terms of lexicographical order, then $\psi_m^0$ represents the 3-mer \A\A\A~, $\psi_m^{1}$ is \A\A\C~, and $\psi_m^{63}$ represents \T\T\T~for all $m$.  For brevity, we also refer to states at event index {\tt m} by their index term {\tt n}.}.  For example, if $\beta\texttt{[n][m{-}1]}=l$ is computed, this means that the state $\psi_m^n$ is part of a path through the trellis who's preceding state is most likely to be $\psi_{m{-}1}^l$.  The calculations behind these estimates are discussed below.  

Before further detailing the Fig.~\ref{f:Viterbi} code, we pause to outline the manner in which it would be expected to operate.  In a typical processing use-case, many event streams will be presented to a detector in parallel.  For example, in existing portable sequencing devices, over 500 parallel channels are capable of simultaneously producing event streams.  Although the length of these streams may vary depending on the DNA sample that produced them, for easier management and without loss of generality, they can be partitioned into equal chunks of length $M$ to present a consistent detection processing load.  Since the origin and location of each such chunk within a given stream is known, splicing these back into complete reads is straightforward\footnote{In contrast, recombining the different streams to form a contiguous genome requires a separate set of algorithms, not considered in this work.}.  More importantly, to avoid the need for 100s of parallel detector instantiations, it is preferred that the code of Fig.~\ref{f:Viterbi} can process channels in a rapid time-multiplexed manner.  Specifically, if there are $C$ channels and each is able to produce data at a rate of $R$ events per second the algorithm in Fig.~\ref{f:Viterbi} has to complete its outer loop in a time of $M/(C{\cdot}R)$ to keep-up, it is this performance pressure that motivates the SoC accelerators discussed in this paper.  

\begin{figure}
\centerline{\includegraphics[width=3.5in]{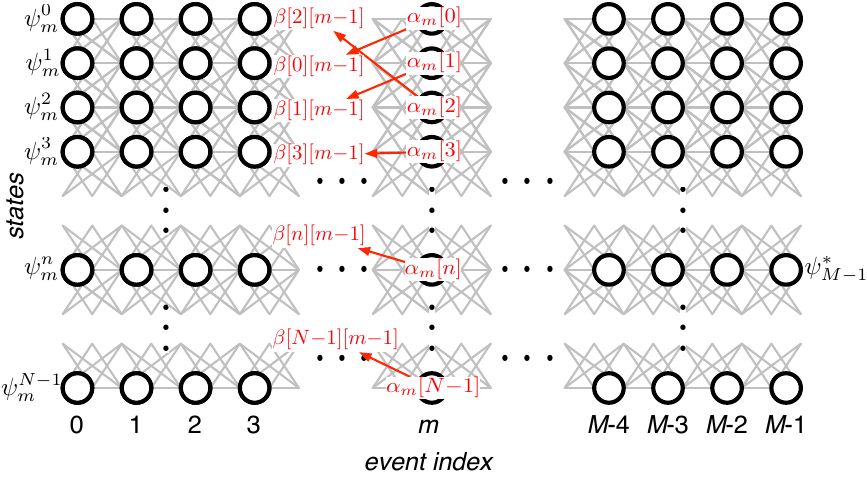}}
  \caption{A trellis representation and its construction via state progression.}
\label{f:trellis23}
\end{figure}

Returning to detailed computational considerations, the algorithm's job is to form probability relationships between events and $k$-mer states.  This is achieved through the construction of a path metric through all the state vectors, $\Psi$, comprising the trellis structure, an action expressible as
\begin{equation}\label{e:path}
p(\bm{x}|\Psi)P(\Psi) = \prod_{m=1}^{M-1} \left[p(x\texttt{[m]}|\Psi_m)P(\Psi_m|\Psi_{m-1})\right]
\end{equation}
where $P(\Psi_m|\Psi_{m-1})$ denotes a probability of {\it transition} between state vectors at adjacent event indexes and where $p(x\texttt{[m]}|\Psi_m)$, denotes the {\it observation} likelihood of a $k$-mer state vector being associated with a given measured event $x\texttt[m]$.  For improved hardware efficiency, the negative logarithm of these terms is computed in the form of the (log) posteriors
\begin{equation}\label{e:alpha}
\{\alpha_m\texttt{[n]}\}_0^{N{-}1} = -\log \left[p(x\texttt{[m]}|\Psi_m)P(\Psi_m|\Psi_{m-1})\right]
\end{equation}
Thus, for each event $x\texttt{[m]}$, the trellis computation updates $N$ terms $\alpha_m\texttt{[0]}$ to $\alpha_m\texttt{[N-1]}$.  For each newly computed posterior $\alpha_m\texttt{[n]}$, a corresponding trellis pointer $\beta\texttt{[n][m-1]}$ is also calculated.  To help clarify, a picture associating these variables to the trellis structure is offered in Fig.~\ref{f:trellis23}. 

The posterior and pointer calculations in the trellis construction algorithm outlined in Fig.~\ref{f:Viterbi} consists of four nested for-loops.  The outermost loop (1), referred to as the {\it event loop} (lines 1–17), sequentially processes incoming events $x\texttt{[m]}$. This prompts the computations of each new set of $N$ states within a state vector $\Psi_m$.  These state calculations are handled within the (2) {\it state loop} (lines 2-11).  Within the state loop, the (3) {\it trans loop} (lines 4-7) computes transition probabilities between states.  After the state loop, a normalization (4) {\it norm loop} (lines 12-16) is run to readjust the computed posterior terms to prevent overflow.

The state loop consist of three main phases.  The first of these is on line 3 of Fig.~\ref{f:Viterbi}, therein the {\tt GatherTrans} function computes and gathers the addresses for all the states at the previous index {\tt m-1} that may transition into the state {\tt n} at the present index {\tt m} (i.e., the current state $\psi_m^n$).  The starting address of the set of states computed by {\tt GatherTrans} is indicated by pointer {\tt transidx}.  In theory, all $N$ preceding states at {\tt m-1} may transition into any one current state at {\tt m}.  This would require computing $N^2$ total transitions for each iteration of the event loop.  But sequence constraints allow only a unique subset of $T{<}N$ states to be considered as transition states into $\psi_m^n$.  Hence, only a total of $N{\cdot}T$ transitions must be identified for each event loop iteration.  Identifying these particular transitions for each state is the job of {\tt GatherTrans}.  

In particular, for any $\psi_m^n$, only $T=21$ preceding states need to be accounted for.  These 21 states reflect so-called {\it stay} (1 in total), {\it step} (4 in total), and {\it skip} (16 in total) transition types.  The stay transition types reflect the possibility that consecutive events at {\tt m} and {\tt m-1} are just separate measurements of exactly the same $k$-mer.  Hence, they simply account for a transition between identical states (i.e., $\psi_{m{-}1}^n$ to $\psi_m^n$).  The step types reflect the possibility that consecutive events reflect 1-mer shifts, for example, like the transition from $\psi_{m{-}1}^6$ to $\psi_m^{27}$ which reflects \A\C\G~``stepping'' to \C\G\T.  And skip types reflect the possibility that only a coarse 2-mer jump has been captured by consecutive events (e.g., \A\C\G~``skipping'' to  \G\A\C, a jump from $\psi_{m{-}1}^6$ to $\psi_m^{33}$).  Generally, the relations between state labels for steps and skips can be as expressed with
\begin{align}
x & = l \cdot 4^{k-1} + \lfloor y/4 \rfloor~\text{for}~\psi^x_{m{-}1}\xrightarrow{\text{step}}\psi^y_{m} \label{e:step_state}\\
x & = L \cdot 4^{k-2} + \lfloor z / 4^2 \rfloor~\text{for}~\psi^x_{m{-}1}\xrightarrow{\text{skip}}\psi^z_{m} \label{e:skip_state}
\end{align}
where $l \in \{0,1,2,3\}$ and $L \in \{0,1,\ldots,4^2-1\}$.  We return to these relations in \S~\ref{ss:design_traceback} when considering hardware design for traceback acceleration.

The second major phase of the state loop concerns execution of its trans loop (lines 4-7) block.  This part computes the (log) probability of the transition term, {\tt trans[:]}=$P(\Psi_m|\Psi_{m-1})$ in equation~\eqref{e:alpha}.  Computationally, it is an application of the probability chain rule expressed as an addition (line 6) between the (log) posterior $\alpha_{m{-}1}$ and the (log) probability of transition between states, {\tt tprob}.  This latter value is determined by some preliminary model training procedure that is outside the scope of this paper.

The third major phase (lines 8-10) of the state loop completes the main trellis construction for event {\tt m} by updating posterior and the corresponding set of trellis pointers for all states.  Specifically, via the {\tt FindMin} function (line 8), this part of the state loop identifies {\tt minidxT}, the most likely of $T$ preceding states to transition into the current state {\tt n}.  This allows the corresponding trellis pointer $\beta${\tt[n][m-1]} to be assigned (line 9).  The new log posterior of state {\tt n}, $\alpha'_m${\tt[n]}, is computed with the {\tt Post} function which effectively completes the calculation summarized in equation~\eqref{e:alpha} as
\begin{align}\label{e:logemission}
\alpha'_m\texttt{[n]} =~&\texttt{trans[minidxT]} - \nonumber \\ 
& \sigma\texttt{[n]} + (x\texttt{[m]} - \mu\texttt{[n]})^2
\end{align}
where $\mu$ and $\sigma$ are another set of model parameters that may be simultaneously determined alongside {\tt tprob} as noted above.

Finally, as noted above, to prevent overflow, a normalization loop (lines 14-15) is executed to produce a scaled set of posteriors $\alpha_m${\tt[n]} to process in the following iteration of {\tt m}.

\subsection{Sequence Detection: Traceback}\label{ss:Traceback}

The Fig.~\ref{f:Viterbi} trellis construction code described above computes all the possible ways in which the event signal $x\texttt[m]$ may be associated with the set of possible $k$-mer states $\Psi$.  With these computations in hand, the most likely monomer output sequence (a DNA read) can be extracted by judiciously selecting a series of trellis pointers $\beta$ to identify the most likely sequence of state through the trellis: $\{\psi_0^*,\ldots,\psi_{M{-}1}^*\}$.  This task is accomplished by the traceback algorithm whose essence is summarized by the pseudocode listed in Fig.~\ref{f:Traceback}.

\begin{figure}
\begin{lstlisting}
state_path = []
prev_state = minidxN
Cat(state_path,minidxN)
for m = M-2 to 0 {  // *(1) traceback loop start
  prev_state = @{$\color{red}{\beta}$}@[prev_state][m]
  Cat(state_path,prev_state)
} // end traceback loop (1)*
\end{lstlisting}
  \caption{The traceback algorithm.}
\label{f:Traceback}
\end{figure}

\begin{figure*}
\centering
\includegraphics[width=7.0in]{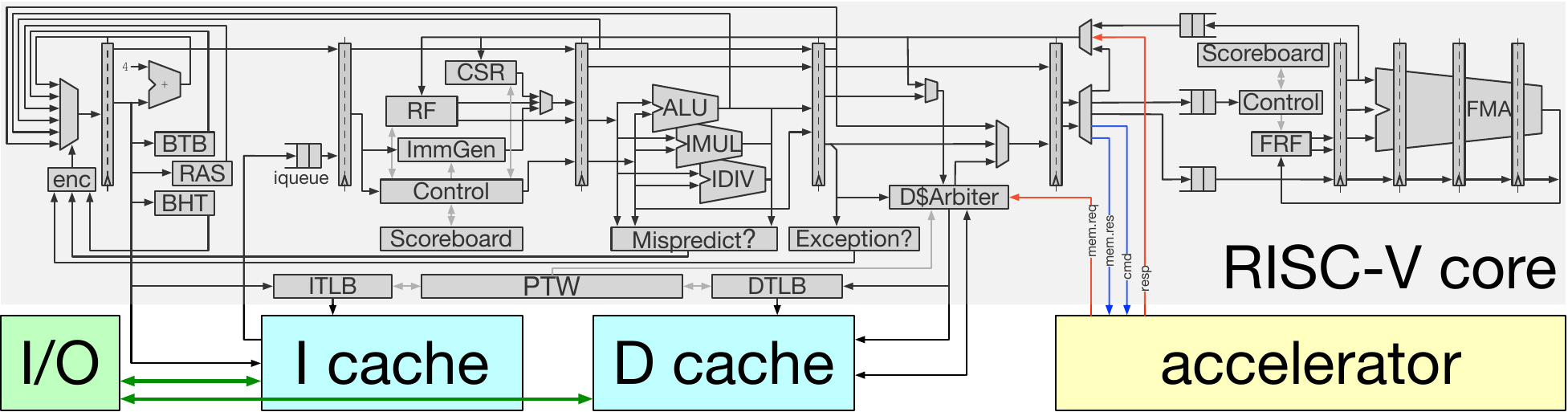} 
  \caption{High-level outline of proposed SoC for next-generation IoT DNA sequencers.  The system consists of a general-purpose RISC-V processing core coupled with on-chip cache memory, application-specific acceleration, and an input/output system containing uncore logic and source-synchronous communication with external components.}
\label{f:ap2}
\end{figure*}

The traceback program begins by initializing the optimal path sequence array, {\tt state\_path}.  From the trellis construction algorithm (Fig.~\ref{f:Viterbi}, line 12), the traceback program grabs the most likely end state, $\texttt{minidxN}=\psi_{M{-}1}^*$.  It then joins this end value with {\tt state\_path} via the concatenation function {\tt Cat}.  The {\it traceback loop} (lines 4-7) then effectively executes a pointer chase through repeated references to $\beta$.  This finds the most likely preceding sequence of $M-1$ states $\texttt{prev\_state}=\psi_m^*$.  As these are identified, they are concatenated with the growing {\tt state\_path} array.

\subsection{Workload Partitioning Considerations\label{ss:hscd}}
Given the specifications (an algorithmic workload) along with the target design space (an embedded RISC processor equipped with specialized hardware accelerators), the miniature sequencer design problem fits squarely within the hardware/software co-design (HSCD) paradigm. In that context, the design problem is centered around answering two key questions - how is the platform selected from the specifications and then how is the application mapped to it. While formal design methodologies exist (particularly for application mapping ~\cite{Eleftheriadis2024}), we employ a coarse design space search to arrive at our solution. Namely, the partitioning strategy for the sequence detection workload previously outlined in Figures ~\ref{f:Viterbi} and ~\ref{f:Traceback} is based on the data-intensive nature of the algorithms and seeks an implementation that minimizes communication across the application-specific integrated circuit (ASIC)-CPU interface. We therefore consider three arrangements for our system which progress from software only and add dedicated hardware in increasing fashion until overall system requirements are met. 

\section{SoC System Outline}\label{s:design}
We now turn our discussion to the SoC system we propose for next-generation IoT DNA sequencers.  As noted in \S~\ref{s:introduction}, our aim is to enhance the computing capability embedded within miniature sequencing devices with added care to limit their power consumption.  Even today, small DNA sequencers can gather a tremendous amount of data in a brief amount of time.  Thus, the ability to compute on this information within the device itself has the potential to greatly reduce the amount of data that needs to be exchanged with external components.  This may therefore appreciably enhance the utility of the IoT network in which such devices are used.

The computational enhancements that may be added to the device are constrained by the desire to accommodate existing or even reduced physical device footprints.  Currently, miniature sequencers are as small as 10${\times}$2${\times}$3~cm$^3$ and contain three main electronic parts - the readout, glue logic, and USB communications as mentioned in \S~\ref{ss:MiniSeqDescription}.  In light of this, limiting any additional computing to a single-chip solution is preferable.  The mobility inherent to such small devices also emphasizes the need to achieve suitable performance within a low power-budget.  Existing devices operate from a 5-W USB supply although likely require less than 2-W for typical operating cases.

With these limits in mind, a high-level depiction of the SoC we are proposing for next-generation IoT DNA sequencers is shown in Fig.~\ref{f:ap2}.  The proposed system centres around a RISC-V core - an open-source Rocket microarchitecture available in the Chipyard repository~\cite{amid2020chipyard}.  Thus, both the system's instruction-set architecture (ISA) and processor microarchitecture are freely available to be used in custom designs, a significant potential cost savings for future developers.  

In particular, the RISC-V core is a 5-stage, in-order, 64-bit implementation fitted with a number of components that, along with the ISA itself, make this architecture capable of supporting a sophisticated OS like Linux.  Although, simpler versions of the core could have been implemented (e.g., a 32-bit architecture), our intention was to initiate exploration on the side of greater general-purpose computing ability.  This would ensure maximal application flexibility while informing future contributions on the power requirements of processors that emphasize performance.  As part of our core's extended capability, we included a floating-point multiply and accumulate unit, prefetching logic, branch prediction plus return address predictors, virtual addressing capability (including translation lookaside buffers), and exception handlers.  To help manage long-latency compute, the architecture also includes scoreboarding logic.  A summary of the processor's key settings is given in Table~\ref{t:soc_params}.

\begin{table}\centering
\caption{\label{t:soc_params}SoC Processor and Accelerator Parameters}
\begin{tabular}{@{}ll@{}}
\toprule
\begin{tabular}[c]{@{}l@{}}Design Parameter\end{tabular} & Value     \\ 
\midrule
\textbf{RISC-V ISA}                                 & RISCV64G \\
\textbf{Data width}                                 & 64 b     \\
\textbf{Instruction Cache} (4-way)                  & 16 KiB   \\
\textbf{Data Cache} (4-way)                         & 16 KiB   \\
\textbf{Cache line size}                            & 64 B     \\
\textbf{Translation lookaside buffer (TLB) entries} & 8        \\
\textbf{Branch history table (BHT) entries} (2-b)   & 4096     \\
\textbf{Branch target buffer (BTB) entries}         & 62       \\
\textbf{Return address stack (RAS) entries}         & 2        \\
\textbf{Accelerator on-chip memory}                 & 72 KiB   \\ 
\bottomrule
\end{tabular}
\end{table}

The processor's on-chip memory comes in the form of a 32-KiB split cache.  This total is divided evenly between the instruction cache (I cache) and data cache (D cache) both of which are set associative with 4-ways and 64~B cache lines.  To help facilitate more efficient program execution and data flow, the D cache is non-blocking with a two-entry miss-status handling register.  

The on-chip memory, as well as the core itself, have access to a high-speed input/output (I/O block in Fig.~\ref{f:ap2}) facilitated through uncore communication logic.  It is through this connection that program and data information is exchanged between the SoC and external sources.  As with the other components, the I/O block is also an open-source design.  This is a multi-stage system that starts with the uncore (from Chipyard) and is followed by a source-synchronous communications (SSC) chain from the Bespoke Silicon Group (BSG) BaseJump project~\cite{BSG18}. The uncore element helps queue core and memory requests to external components via the TileLink protocol~\cite{Cook17} while the SSC component implements back-pressure and flow-control to help manage chip-to-chip communications.

Finally, as indicated in Fig.~\ref{f:ap2}, our SoC includes bioinformatics hardware accelerator blocks that implement the trellis construction and traceback codes given in Figs.~\ref{f:Viterbi} and~\ref{f:Traceback}, respectively.  By exploiting parallel processing opportunities these designs significantly improve performance over processor-only implementations.  By replacing expensive RISC instruction fetches and maintaining localized data movement within the accelerators, these performance gains are achieved alongside substantial improvements in energy efficiency.  We detail the design of this unit in the following section.

\section{Accelerator System Design}\label{s:accelerator}

\begin{figure}
\centerline{\includegraphics[width=3.5in]{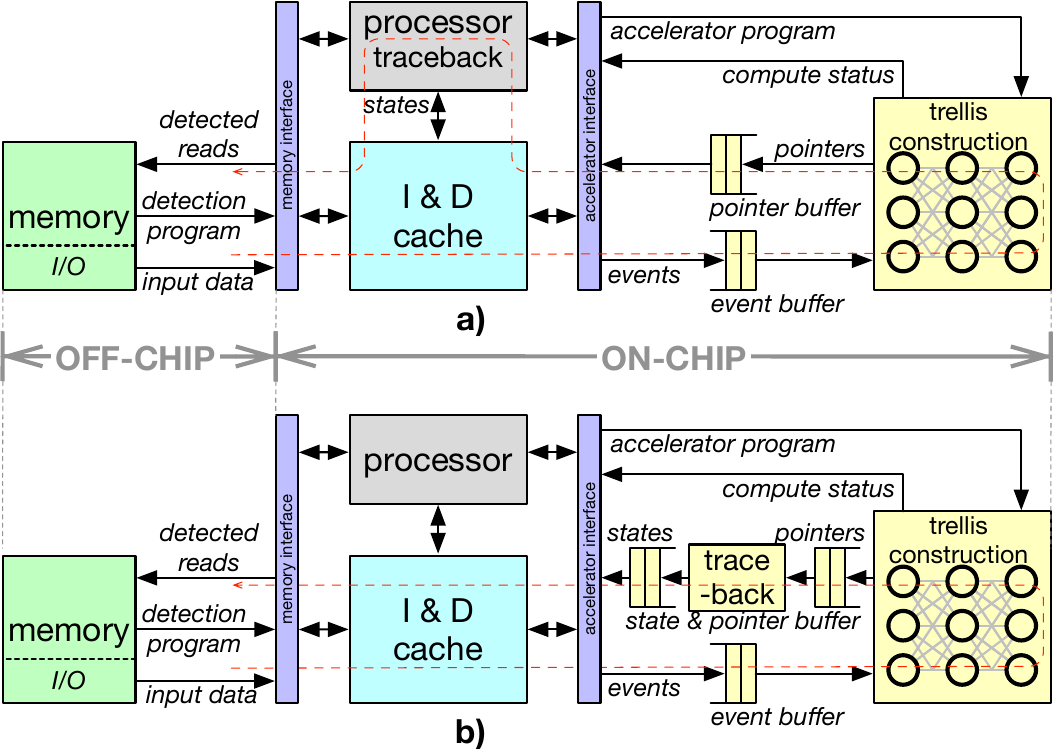}}
  \caption{System outline of two accelerator designs realized in this work.  a) AccelA: an accelerator for trellis construction only; b) AccelB: an accelerator merging both trellis construction and traceback functions.}
\label{f:accelversions}
\end{figure}

A system-level outline of the accelerators implemented in this work is given in Fig.~\ref{f:accelversions}.  On-chip, two accelerator versions, AccelA and AccelB, are attached to a single core (for clarity, the accelerators are shown separately in Fig.~\ref{f:accelversions}).  AccelA, tasked only with processing the trellis construction algorithm of Fig.~\ref{f:Viterbi}, is made to work in tandem with the core (which computes the traceback code itself).  We have previously discussed this acceleration style in~\cite{Dawji2024}.  AccelB employs the same trellis constructor implementation, but adjoins it directly to hardware tasked with executing the traceback code of Fig.~\ref{f:Traceback}.  As shown in the chip measurement discussions of \S~\ref{s:measurement} this addition brings substantial benefits to the SoC's performance.

The general operation of AccelA and AccelB is similar.  Both processors receive a series of accelerator program commands from the core. These commands set the number, $M$, of events $\bm{x}$ to process and the location of the events in memory.  The memory locations of model parameters {\tt tprob}, {\tt mu} and {\tt sigma} are also conveyed as is the memory location for data returned by the accelerator.  Once the program is loaded the accelerator commences operation.

In the case of AccelA, this involves feeding in events from memory through a 4-KiB SRAM {\it 
 event buffer} and returning trellis pointers to memory through a 32-KiB SRAM {\it pointer buffer}.  Event loading and pointer return to/from buffers can be initiated simultaneously.  After all events are processed, AccelA informs the core of its completed compute status and the traceback algorithm is subsequently computed by the processor.  The computations on another set of $M$ pointers may then commence.

AccelB operates in a manner similar to the AccelA method outlined above.  It differs in that, after a set of trellis pointers is computed by the trellis constructor hardware, these are then sent to a traceback hardware block that computes the corresponding state sequence.  This sequence is then sent back to memory along with a compute status signal to the RISC-V core.  This status signal enables the core software to initiate another set of events for processing by the accelerator.  More details on the manner in which signals are exchanged between the core and the accelerator are provided next.

\subsection{Accelerator Software Interface}\label{ss:interface}

The accelerators used in this work both connect to the core and its D cache using the Rocket Custom Coprocessor (RoCC) interface, another open-source Chipyard component.  RoCC facilitates 64-bit communication between accelerators and the processor using sets of request/response channels.  As shown in Fig.~\ref{f:ap2}, four RoCC ports are used in this work, {\it cmd} and {\it resp} for control and status signal exchange between the RISC-V core and the accelerators and {\it mem.req} and {\it mem.resp} for data load/store between the D cache and the accelerators.  The RoCC ports support 64-bit data paths to and from memory as well as 40-bit memory address ports.

Software communicates with the accelerator through extended 32-bit assembly instructions that map directly to instructions in the RISC-V ISA designed to forward commands to the RoCC logic and custom accelerator hardware.  As a result, commands such as

\begin{footnotesize}
\begin{verbatim}
asm volatile ("custom0 %[rd],%[rs1],%[rs2],0" : \
[rd]"=r"(z) : [rs1]"r"(x), [rs2]"r"(y)); 
\end{verbatim}
\end{footnotesize}
can be included in developer C code.  Thus, contents in the RISC-V core's register file locations, {\tt x} and {\tt y} may be sent to the accelerator (with results written back from the accelerator to a core register {\tt z}).  Wrapping such commands in C macros allows their low-level details to be hidden from software developers.

In this work, the accelerators under study could be operated using a series of 6 macro commands (the aforementioned accelerator program) that form the following programming sequence: 1) accelerator reset; 2) set number of events, $M$, to process; 3-5) starting memory addresses for model parameters; 6) starting memory address for events $x${\tt[0]} and starting memory address for computational results.  With the successful receipt of the sixth command the accelerators initiate hardware-based execution of the algorithms in Figs.~\ref{f:Viterbi} and~\ref{f:Traceback}.  This process is conducted between the accelerators and memory units (cache and DRAM) until all the events are processed and associated outputs are generated.  Since the RISC-V processor uses virtual addressing, it conveniently integrates the accelerator's load/store actions within the standard memory space.  Thus, programming for the exchange of data between the processor's main memory and the accelerator is straightforward.    

\subsection{Trellis Constructor Hardware Design}\label{ss:design_trellis}

The accelerator processes loop (1) (the event loop) of Fig.~\ref{f:Viterbi} program sequentially.  The serial nature of the input data and of the algorithm enforce this constraint since calculations on state vectors $\Psi_m$ depend on previous results $\Psi_{m{-}1}$.  As discussed in \S~\ref{ss:Trellis}, executing this loop at a sufficiently high rate, $R$, affords the prospect of realizing effective parallel processing on $C$ buffered event input channels.

To help maximize the achievable $R$, the accelerator hardware implements fully unrolled versions of loops (2)-(4).  This pertains to the independence of the calculations within these loops.  Thus, loops (2) and (3), the state and trans loops, can be largely executed in parallel followed by the execution of unrolled loop (4), the norm loop.  The size of these inner loops depend on $N$ and $T$.  As explained in \S~\ref{ss:Traceback}, the $T$ setting is fixed at 21 for this algorithm and, for our implementation, we chose $N{=}64$, a reflection of the size selected for a DNA detector discussed in~\cite{Timp12}.  The system is designed to compute a maximum of $M{=}512$ input events at-a-time.  As discussed in~\S~\ref{ss:Trellis}, should DNA samples exceeding this length need to be processed, their measured event sequences would be split into $M{\le}512$ long chunks by the RISC-V core for accelerated processing.  The detected chunks would then be reconstituted into their final read form.

\begin{figure}
\centerline{\includegraphics[width=3.5in]{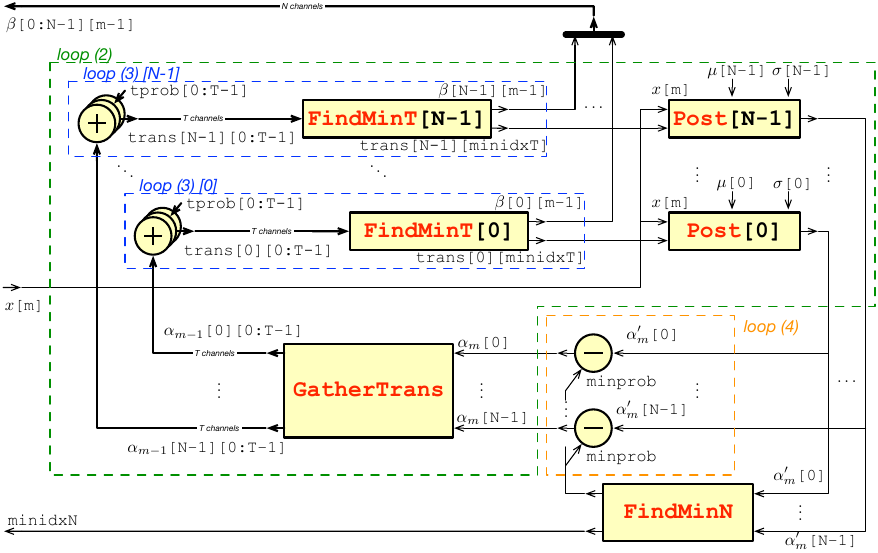}}
  \caption{The arrangement of the trellis construction hardware accelerator and its relation to the algorithm in Fig.~\ref{f:Viterbi}.}
\label{f:trellishw}
\end{figure}

The hardware arrangement of the trellis construction accelerator is shown in Fig.~\ref{f:trellishw}.  The organization of its component parts maps closely to the algorithm description in Fig.~\ref{f:Viterbi} and the majority of these are dedicated to the execution of loops (2) and (3).  

As with the trellis construction algorithm, the accelerator's signal flow path in Fig.~\ref{f:trellishw} proceeds from the {\tt GatherTrans} block, into a set of $N$ unrolled loop (3) implementations.  The {\tt GatherTrans} block retains the normalized $N$ posteriors computed in the previous iteration of loop (1) (i.e., iteration {\tt m-1}) and provides these as $T$-sized input bundles, $\alpha_{m{-}1}${\tt[n][0:T-1]}, to corresponding loop (3) datapaths.  Since the destinations of {\tt GatherTrans} outputs are fixed, this block is realized as a hardwired posterior redistribution network.

Each of the $N$ loop (3) implementations consist of $T$ adders and a $T$-input {\tt FindMinT[n]} block, the latter being a hardware realization of the {\tt FindMin(trans,$T$)} function in line 8 of Fig.~\ref{f:Viterbi}.  Using the $\alpha_{m{-}1}$ inputs from {\tt GatherTrans}, the adders collectively compute the unrolled version of line 6 of the Fig.~\ref{f:Viterbi} algorithm.  These results are then passed to the {\tt FindMinT} components from which $N$ trellis pointers $\beta${\tt [0:N-1][m-1]} emerge.  The trellis pointers are either passed back to the RISC-V core's memory, as with the AccelA arrangement outlined in Fig.~\ref{f:accelversions}a), or, as with the AccelB design in Fig.~\ref{f:accelversions}b), are buffered and passed to the traceback hardware unit discussed in \S~\ref{ss:design_traceback}.

As shown in Fig.~\ref{f:trellishw}, in parallel with the trellis pointer outputs, the results from all $N$ loop (3) blocks also include the transition probability values {\tt trans[n][minidxT]}.  As discussed above, these results along with the event inputs $x${\tt [m]} are processed by hardware versions of the {\tt Post} to compute the posterior update equation~\eqref{e:logemission}.

Together, these $N$ simultaneous calculations complete loop (2) and are subsequently processed by an unrolled version of the normalization code in loop (4) which subtracts out the minimum posterior calculation.  In-turn, this minimum is computed in Fig.~\ref{f:trellishw} with the {\tt FindMinN} block, a hardware version of the {\tt FindMin($\alpha'_m$,$N$)} function in line 12 of Fig.~\ref{f:Viterbi}.  To initiate the traceback process, the {\tt FindMinN} block also outputs {\tt minidxN}, the most likely end state of the sequence.

A single loop (1) iteration is completed in 18 cycles across the design pictured in Fig.~\ref{f:trellishw}.  Seven of these cycles are spent in loop (3) and the adjoining {\tt Post} blocks.  These calculation result in the completion of a new set of trellis pointers $\beta$ and posterior terms $\alpha'$.  Another seven cycles is consumed by the normalization operation, a job completed by loop (4) and associated {\tt FindMinN} block.  The remaining four cycles are used for communications and handshaking overhead between components internal to AccelA as well as to the buffers managing $x${\tt [m]} inputs and $\beta$ outputs.

\begin{figure}
\centerline{\includegraphics[width=3.5in]{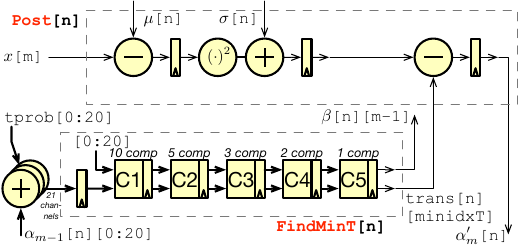}}
  \caption{Architecture of {\tt FindMinT} and {\tt Post} blocks.}
\label{f:transpost}
\end{figure}

A closer look at {\tt FindMinT} and {\tt Post} hardware blocks is provided in Fig.~\ref{f:transpost}.  As per line 6 of the trellis construction algorithm in Fig.~\ref{f:Viterbi}, a group of 21 adders sum 21 posteriors, $\alpha_{m{-}1}$, and associated transition model probabilities, {\tt tprob} to form 21 {\tt trans} terms.  The minimum of these 21 terms, {\tt trans[n][minidxT]} is derived over the course of five cycles using pipelined comparator banks (C1 to C5 in Fig.~\ref{f:transpost}) arranged as a parallel reduction tree.  Besides {\tt trans}, a corresponding set of 21 index values (numbered 0 to 20) are processed by the pipelined comparator.  These effectively track the state location of each trans input {\it relative} to the loop (3) element, {\tt n}, that's processing them.  Thus, by the time a minimum {\tt trans} value is produced by C5, its corresponding index is also output and serves as the corresponding relative trellis pointer $\beta$.  Avoiding global pointer calculations simplifies the accelerator’s hardware, but increases the complexity of the subsequent traceback function, which must resolve states based on their global labels.  We return to this issue in \S~\ref{ss:design_traceback} when describing the traceback acceleration hardware. 

In parallel with {\tt FindMinT}, the operations of the {\tt Post} block are executed as also shown in Fig.~\ref{f:transpost}.  This is another pipeline structure that completes the updated posterior calculation shown in equation~\eqref{e:logemission} by combining event ($x$) and model ($\mu$ and $\sigma$) inputs with the {\tt trans} result from {\tt FindMinT}.  The output from {\tt Post} is normalized as shown in Fig.~\ref{f:trellishw}.  In particular, the organization of {\tt FindMinN} is nearly identical to that of {\tt FindMinT} except that six comparator stages are used to find the minimum from 64 {\tt Post} outputs.

\subsection{Traceback Hardware Design}\label{ss:design_traceback}

\begin{figure}
\centerline{\includegraphics[width=3.5in]{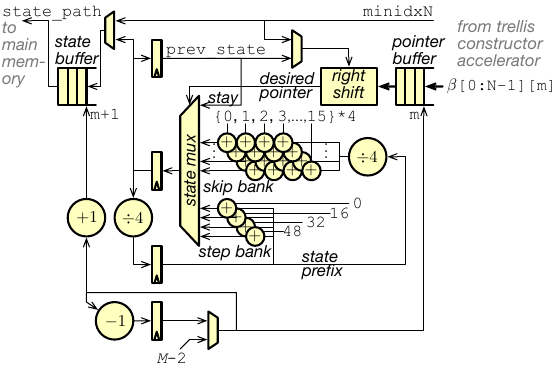}}
  \caption{Architecture of the traceback acceleration hardware.}
\label{f:traceback}
\end{figure}

As noted earlier, AccelB includes not only an engine for trellis pointer calculations, but augments it with hardware to accelerate the traceback function as well.  A sketch of the architecture of this portion of the accelerator is shown in Fig.~\ref{f:traceback}.  As outlined in Fig.~\ref{f:accelversions}b) above, the traceback computations are located between two buffers: a {\it pointer buffer} at the input and a {\it state buffer} at the output.  The trellis buffer, an 8-bank 32-KiB SRAM, accumulates trellis pointer outputs, $\beta$, from the trellis construction engine; the state buffer, a 384-B register file, accumulates the final state sequence derived by the traceback computations on the trellis pointers.  When completed, this state sequence is streamed out of the state buffer to the RISC-V processor's memory (via the RoCC interface discussed in \S~\ref{ss:interface}, but not shown in Fig.~\ref{f:traceback}) as the {\tt state\_path} term present in the algorithm listed in Fig.~\ref{f:traceback}.

The traceback hardware's computational units wait until all pointers corresponding to an $M$-long event sequence are accumulated in the pointer buffer.  Once, a control line (not shown in Fig.~\ref{f:traceback}) from the trellis constructor hardware signals the delivery of all trellis pointers, traceback's state sequence calculations commence for $M{-}1$ iterations of the traceback loop (1) highlighted in Fig.~\ref{f:traceback}.  

At each iteration the traceback mechanism grabs, from the pointer buffer, a corresponding string of $N$ relative trellis pointers computed for the event corresponding to that iteration.  Specifically, this $N$-pointer string is read from the pointer buffer as selected by the address signal {\tt m}.  As in the algorithm of Fig.~\ref{f:traceback}, this fetch is done in a last-in-first-out fashion starting at the pointer buffer's location {\tt m}$=M{-}2$.  From this fetched string, the {\it desired pointer} is chosen by the {\tt prev\_state} term and a right shift block.  In particular, the desired trellis pointer is extracted by shifting the fetched pointer string to the right by {\tt prev\_state} bytes.  Thus, the least-significant byte out of the right shift is the desired trellis pointer.  As in the traceback code of Fig.~\ref{f:traceback}, at the start of execution, the desired pointer is selected by the {\tt minidxN} value provided by the trellis construction accelerator.

As noted in \S~\ref{ss:design_trellis} the trellis pointers provided by the trellis constructor accelerator are relative to the unrolled loop component that generated them.  In particular, they just enumerate the 21 possible transitions from one state to another in terms of the numbers 0 to 20.  To construct a proper {\tt state\_path} however, these values must be converted to the global state index range that spans 0 to 63 in this application.  To re-compose these relative trellis pointers into their global equivalents, the expressions~\eqref{e:step_state} and~\eqref{e:skip_state} discussed in \S~\ref{ss:Trellis} can be applied (where $y$={\tt prev\_state}).  These expressions are implemented by the arithmetic adder banks and ${\div}4$ blocks (2-bit shifters) in Fig.~\ref{f:traceback}.  

Specifically, at each new iteration {\tt m}, the previously computed global state, {\tt prev\_state} is fed back directly (where it represents the stay transition) as well as through various ${\div}4$ blocks and adder banks.  These paths generate the 21 possible global states which may transition into {\tt prev\_state}.  The outputs of these paths are connected to a {\it state multiplexer} (see Fig.~\ref{f:traceback}) whose select port is driven by newly fetched relative trellis pointers.  With relative trellis pointers corresponding to the desired stay, step, or skip transitions, thus, the desired global state values are computed at each stage and queued onto the state buffer.

\section{Implementation and Measurements}\label{s:measurement}
\begin{figure}
\centerline{\includegraphics[width=2.5in]{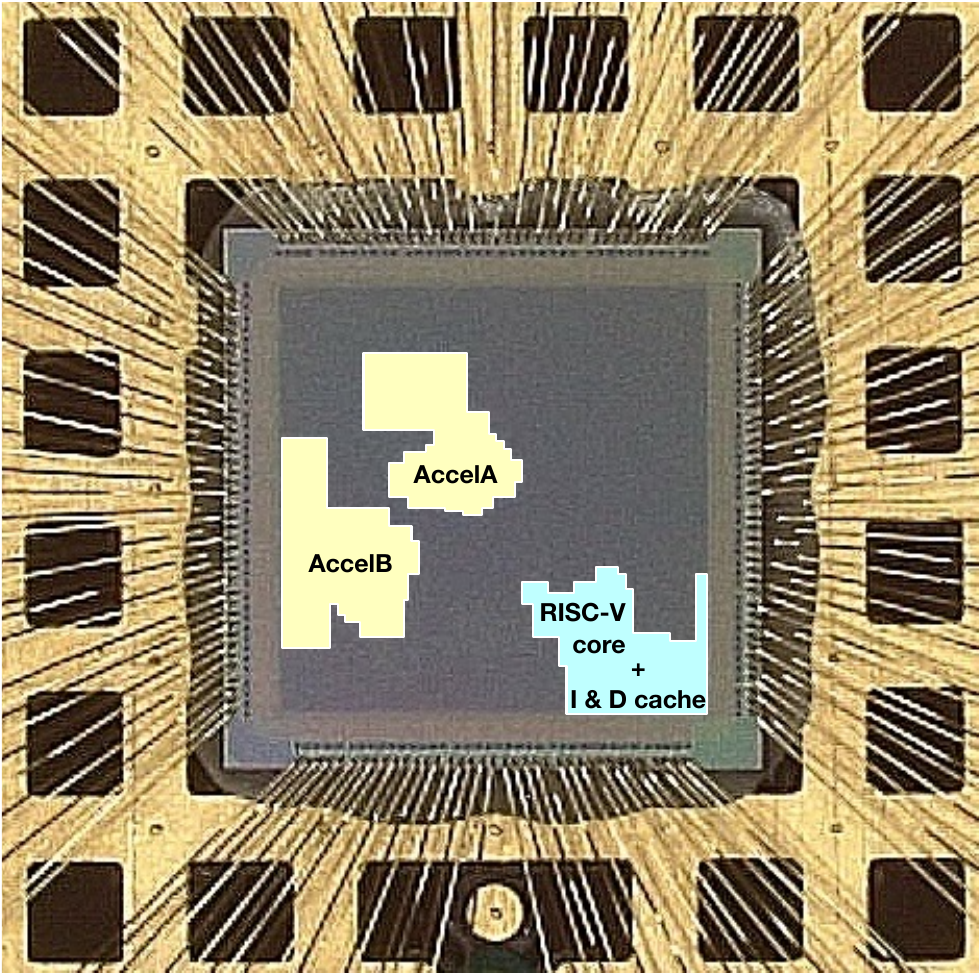}}
  \caption{Mobile DNA sequencing system-on-chip bonded die photo.}
\label{f:chip}
\end{figure}

The sequence detection SoC described above is implemented in \textsc{GlobalFoundries} 22-nm fully-depleted system-on-insulator (FDSOI) CMOS process.  A photo of the bonded chip is shown in Fig.~\ref{f:chip}.  The chip is wirebonded to a ball grid array (BGA) package substrate and placed within an elastomer contact BGA socket.  The footprints of the main hardware components of this SoC -- the RISC-V core and its cache, AccelA, and AccelB -- are also highlighted in the picture.

The microprocessor and accelerators are synthesized from Verilog and Chisel descriptions of the hardware.  The chip area occupied by the core, AccelA, and AccelB are 0.355, 0.395 and 0.401~mm$^2$ respectively.  About 30\% of the RISC-V core's area is consumed by its logic and register blocks, the remainder being taken up by the D and I cache components.  The logic contribution to area in the accelerators is higher - about 45\% for both - with the remainder consumed predominantly by the event and pointer buffer memories.
\subsection{Test Platform}
\begin{figure}
\centerline{\includegraphics[width=3.5in]{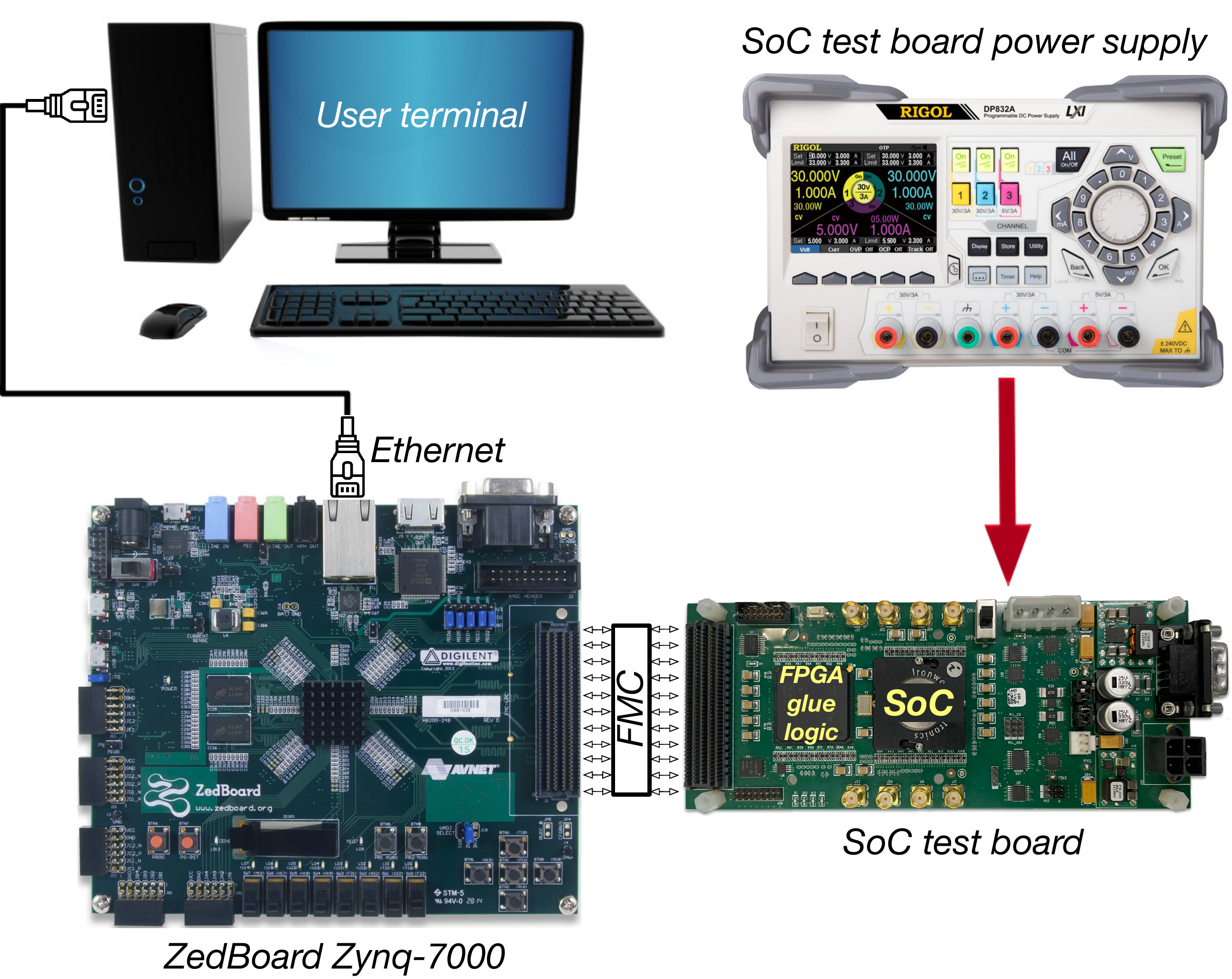}}
  \caption{The packaged SoC within its test system.}
\label{f:rtboard}
\end{figure}

Fig.~\ref{f:rtboard} shows a picture of the packaged SoC implemented as part of an open source board-level test system originally developed and described by the BSG group~\cite{BaseJump23}.  As shown, the hardware test setup consists of two boards, a custom SoC test board to which the packaged SoC is affixed and a ZedBoard Zynq-7000 ARM/FPGA SoC Development Board (``ZedBoard'').  The two boards are attached via a low-pin-count FPGA mezzanine connector (LPC FMC).  A Spartan-6 FPGA shares the test board with the SoC and serves to provide, clocking, measurement, and interfacing functions.  Although intended as a test facilitator, as mentioned in \S~\ref{ss:MiniSeqDescription}, the glue logic provided by the test board's adjoining FPGA reflects the presence of such devices in contemporary miniature DNA sequencers.  

The adjoining ZedBoard provides the SoC with external memory and a convenient means for users to interface with its RISC-V core via keyboard and a terminal.  Specifically, 256~MiB of the ZedBoard's 512~MiB DRAM are incorporated into the SoCs main memory space with communications between these blocks managed by a hardened DDR memory controller within the Zynq-7000 FPGA.  Memory commands between the daughterboard and ZedBoard are packaged within the AXI protocol.  Also, a separate AXI channel between our chip and the ZedBoard routes a subset of system commands from the SoC's RISC-V core to an ARM Cortex-A9 processor included in the Zynq-7000.  Through an Ethernet connection to an adjoining PC user terminal, the ZedBoard's ARM processor forwards these messages to a shell which allows users to interact with the SoC and its file system.  Via this interface, users can launch code executables, compiled using the RISC-V toolchain, on the SoC. 

\subsection{Functional Verification}
\begin{figure}
\centerline{\includegraphics[width=3.5in]{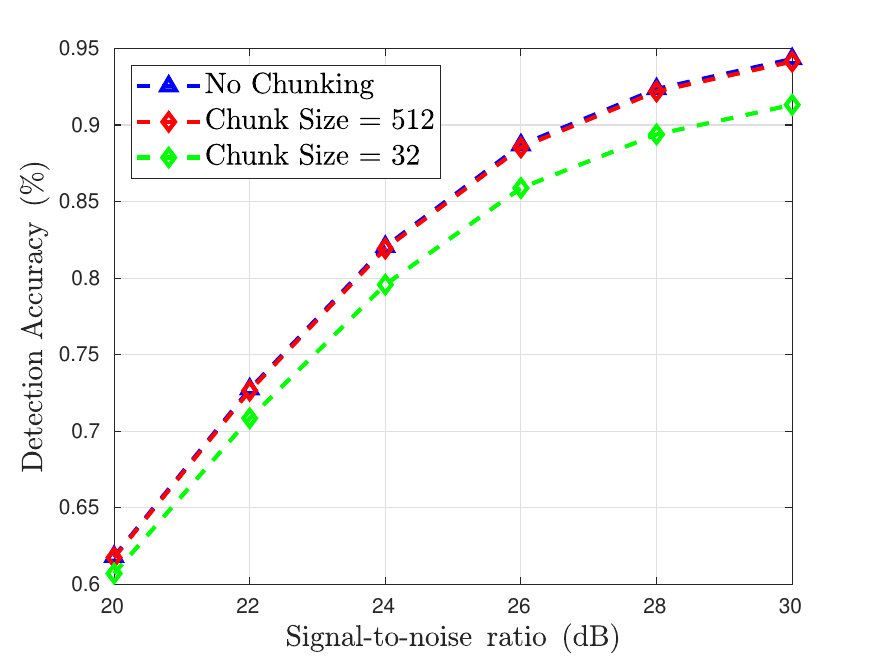}}
  \caption{Accuracy as a function of chunk size}
\label{f:accuracy}
\end{figure}

Our evaluation of the hardware begins with a functional assessment. As previously mentioned, the fixed size of the accelerator buffers requires breaking up and processing the input event sequence in M sized chunks. To determine the impact on accuracy as well as verify correctness with respect to the reference software, detection is performed on a data set of 1800 Matlab-generated base sequences for which predictive nanopore 3-MER models are used for event generation. The results are illustrated in Fig.~\ref{f:accuracy} which shows detection accuracy as a function of the nanopore signal-to-noise ratio (i.e the noise level in an individual event measurement). The blue curve represents the ideal case where the algorithm is applied to the input sequence in its entirety without any chunking performed. The red and green curves represent event sequences processed in chunk sizes of 512 and 32 events respectively and report the accuracy of the assembled output. Each point is the average accuracy of 100 input samples. As the curves show, detection via chunk sizes as low as 32 still yields above 90\% accuracy. Furthermore, the hardware’s output matches the reference software exactly, confirming functional correctness.

\subsection{Performance Evaluation}
For performance evaluation and comparison we measure the event rate $R$, discussed in \S~\ref{ss:Trellis}.  This refers to the speed at which our SoC can process streams of events, $x$[m], for the purpose of computing their corresponding DNA monomer state sequence components, {\tt state\_path[m]}.  That is, we evaluate our performance on the combined execution of the trellis construction and traceback algorithms in Figs.~\ref{f:Viterbi} and~\ref{f:Traceback} within one program.  The algorithms are implemented in C and executed on a variety of hardware besides our accelerated SoC.  The C code is identical for all the platforms examined save for functions that are replaced by our SoC's accelerator hardware equivalents via the software interface described in \S~\ref{ss:interface}.  

Besides examining execution on our SoC using various workload partitioning schemes (RISC-V core only, core+AccelA, core+AccelB), we also measure performance on three other platforms: the Tensilica Xtensa LX6 RISC-V microprocessor within the ESP32 microcontroller clocked at 240 MHz (40-nm CMOS); the Intel Core i5-12400 processor clocked at 4.4 GHz (10-nm FinFet); and the ARM Cortex-A53 MPCore within a Xilinx Zynq UltraScale+ MPSoC ZCU106 Evaluation Kit (16-nm CMOS).  Together these platforms span three different ISAs: RISC-V, x86, and ARMv8.  They also offer a glimpse at execution across three different contexts: cheap, off-the-shelf microcontrollers (ESP-32), desktop (Core i5), and high-end embedded (Cortex-A53).  ISA-specific GNU toolchains are used to target our code for each of these platforms: the RISC-V toolchain v. 5.3.0; the x86 toolchain v. 11.4.0; the Xilinx adapted ARM toolchain v. 11.2.0. For all comparison platforms, our tests show results on code compiled with full optimization settings (use of the catch-all -march=native -O3 flags).  Moreover, our tests exploit each comparison platform's own hardware enhancements to achieve its maximum throughput. For the x86, this meant multi-threaded execution across all CPU cores. For the Cortex-A53, in addition to the parallel CPUs, the NEON single-instruction multiple-data (SIMD) architecture was targeted through the appropriate compiler settings.

\begin{figure}
\centerline{\includegraphics[width=3.8in]{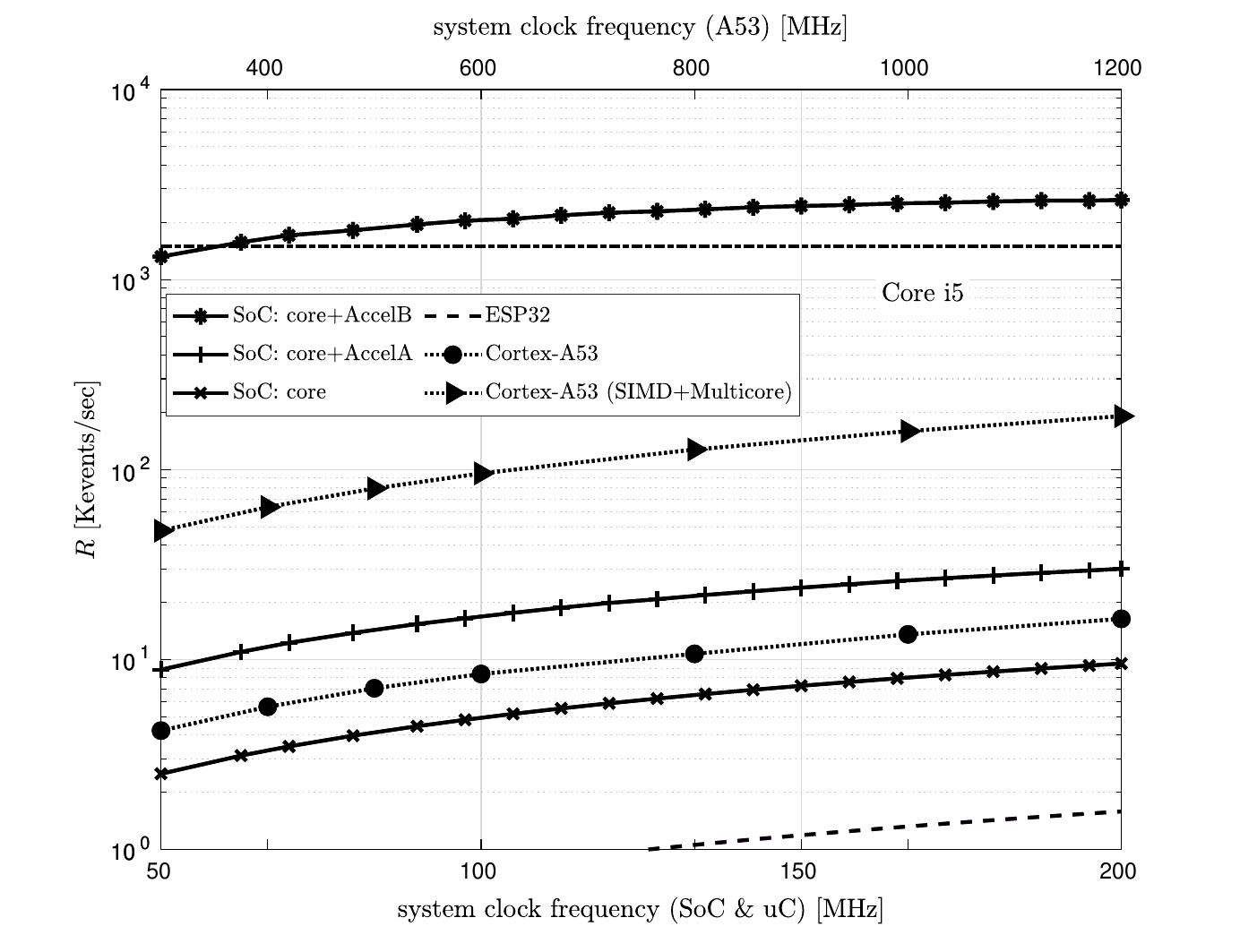}}
  \caption{Measured DNA sequence detection performance comparison between the various SoC configurations and commercial chips representative of desktop and embedded implementations.}
\label{f:speed}
\end{figure}

Our performance measurements are shown in Fig.~\ref{f:speed} as a function of the system clock frequency.  In the case of our SoC (and the ESP32 Microcontroller) this is denoted by the bottom x-axis which goes up to 200-MHz, the maximum achievable processor clock for our 22-nm design.  Since we were only able to control the clock frequency of the Cortex-A53 between 0.3 and 1.2~GHz, this sweep is tracked by the top x-axis labels.  Since we had no ability to control the frequency of the x86 desktop, its performance data point (at 4.4 GHz clock) is represented by the labelled straight line in the figure.

To facilitate a convenient comparison, the performance is displayed on a log-scale.  Although muted by the display scale, as expected, the rate at which we can execute improves linearly with frequency.  At the peak achievable operating frequency of our SoC's core+AccelB arrangement we achieved $R{=}2.6$~Mevents per second.  Across the totality of their parallel channels, modern miniature DNA sequencing devices can measure at rates of roughly 0.2~Mevents per second~\cite{Wick2018}.  Therefore, this version of our SoC operates well clear of present performance expectations and is nicely positioned to support future improvements in DNA sequencers or to sacrifice detection performance for the addition of more functional features.

As is also clear from Fig.~\ref{f:speed}, the core+AccelB configuration of the proposed SoC exhibits a dominant performance advantage over the other systems examined in this work.  Specifically, compared to its counterpart core+AccelA SoC implementation and the core only, the core+AccelB setup possesses a speed-up factor of 87$\times$ and 260$\times$, respectively.  Naturally, the fact that AccelB accelerates multiple functions from the DNA sequence detection program suggests that a performance improvement should be expected. 

The extent of the improvement is driven by a combination of aggressive parallelism and memory proximity.  In particular, with AccelB's trellis construction block we carry out up to {1,300} executions in parallel (via the unrolling of loops (2) and (3) for instance).  In the process of this computation, the accelerator hardware internalizes processing/memory exchanges that would otherwise require about {8,000} memory swaps per event if done with a scalar core alone.  Such work is encapsulated by AccelA and, as shown in Fig.~\ref{f:speed}, it can provide about a 3${\times}$ performance boost compared to the SoC's core only configuration.

Although this boost is helpful, especially when considered in the light of its energy efficiency (discussed below), the potential improvements of the AccelA trellis constructor are throttled by the needs of traceback.  This arises since, as expressed in Fig.~\ref{f:Traceback}, traceback cannot begin until a sufficient set of trellis pointers, $\beta$, have been computed.  If the SoC's core is expected to perform this operation (as in the case of core+AccelA and core only settings), the system is left with two challenges: i) sending the pointer data through a more narrow communications channel between the accelerator and memory and ii) dealing with limited cache capacity.  In our case, the RoCC offers an 8-B communication link between accelerators and memory, but each event generates 64-B worth of trellis pointers.  More importantly, a fully buffered set of 512 events, as possible with our accelerators, will generate 32-KiB worth of trellis pointers.  This easily overwhelms the SoC's 16-KiB D cache.  Thus, despite the relative simplicity of the traceback algorithm (a two-dimensional array traversal), the aforementioned challenges (including accelerated local-to-global trellis pointer calculation) called on dedicated hardware nonetheless and so we integrated traceback acceleration hardware into the trellis construction engine. That allowed our SoC to buffer and traverse pointer results without the need to engage main memory.

Fig.~\ref{f:speed} also includes results for the four other systems noted above: x86, ARM with and without multithreaded, SIMD compilation and the ESP-32. Compared to these, our 200-MHz core+AccelB SoC offers performance boosts of about 1.75$\times$, 13.5$\times$, 163$\times$, and 1300$\times$ respectively.  These comparisons are made to each platform's maximum performance value. The win ultimately stems from the same advantages noted above - mainly in this SoC mode we invoke specialized hardware to perform unrolled trellis computations on the detection code {\it and} couple it with internal traceback acceleration that prevents flooding the core's cache.  An SoC setup that accelerates only the trellis - core+AccelA - even with its heavy emphasis on loop unrolling, is exceeded in performance even by the single-core, SIMD-enabled embedded processor.  Of course, the more specialized nature of our proposed system does offer substantial power savings, an issue we turn to next.

The nominal supply voltage for our computer's 22-nm circuits is 0.8~V, but the SoC and accelerators can operate from levels as low as 0.49~V when working from a 50-MHz system clock and 0.64~V when working with a 200-MHz clock.  Under such reduced voltage circumstances, our highest performance configuration - the core and AccelB working together - consumes a total power of 3.8~mW from the minimal supply at a 50~MHz clock and 20.2 mW when working from a 200~MHz clock. Within these totals, the overhead power of actually running the DNA detection software is 0.5~mW and 2.6~mW at 50 and 200-MHz clock frequencies, respectively.  The remaining power is consumed by the clock when the system is otherwise idle.

\begin{figure}
\centerline{\includegraphics[width=3.8in]{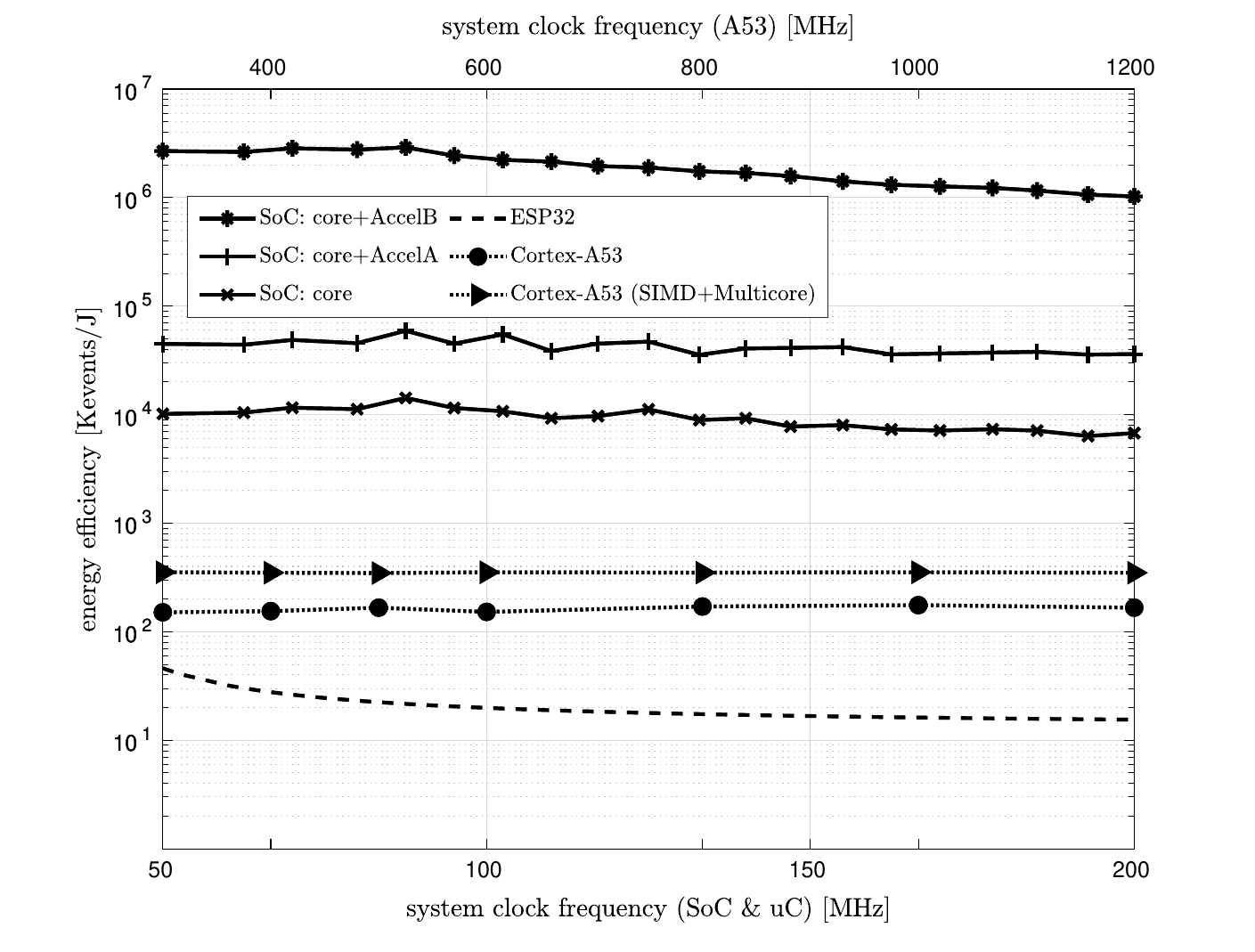}}
  \caption{Measured DNA sequence detection energy efficiency comparison between the various SoC configurations and commercial chips representative of desktop and embedded implementations.}
\label{f:EE}
\end{figure}

A deeper analysis of our system's resource needs is presented in the energy-efficiency plot given in Fig.~\ref{f:EE}. Specifically, over a range of system clock frequencies, we record the number of events that can be processed for a 1 Joule consumption of overhead energy (derived from the aforementioned overhead power).  We do this for various configurations of the SoC as well as the desktop and embedded platforms described above. The SoC's power consumption is obtained directly from the benchtop supply channel feeding the SoC's 0.8V domain. Power consumption for the ARM Cortex is similarly obtained from surrounding board-level hardware responsible for monitoring various power rails. The x86 power is obtained using Intel's Running Average Power Limit Model Specific Registers (RAPL MSRs) - an interface accessible through Linux C API's. Finally, power consumption for the ESP32 is also measured directly from its supply.

\begin{table*}\centering
\caption{\label{t:comp}Performance and Energy Efficiency Comparison Between Platforms}
\ra{1.1}
\small
\begin{tabular}{@{}lccrrr@{}}
\hline
\textbf{Platform} & \textbf{\begin{tabular}[c]{@{}c@{}}Node\\ {[}nm{]}\end{tabular}} & \textbf{\begin{tabular}[c]{@{}c@{}}Processing \\ System\end{tabular}} & \multicolumn{1}{c}{\textbf{\begin{tabular}[c]{@{}c@{}}Perf.\\ {[}Kevents/s{]}\end{tabular}}} & \textbf{\begin{tabular}[c]{@{}r@{}}Energy Eff.\\ {[}Kevents/J{]}\end{tabular}} & \multicolumn{1}{c}{\textbf{\begin{tabular}[c]{@{}c@{}}Energy Eff.\\ Boost Factor\end{tabular}}}\\ \hline
This work (SoC)                        &22&RISC-V+AccelB& 2600&1024$\times$10$^3$&  -- \\
This work (SoC)                        &22&RISC-V+AccelA&   30&  37$\times$10$^3$&   28\\
This work (SoC)                        &22&RISC-V core  &   10& 7000  &             146\\
ZCU106 Embedded ARM                    &16&Cortex-A53   &   16&  170             & 6024\\
ZCU106 Embedded ARM (SIMD + Multicore) &16&Cortex-A53   &   191& 349             & 2934\\
Desktop computer                       &10&Core i5-12400 & 1490&  47 & 22$\times$10$^3$\\ 
Accelerated desktop~\cite{Wu20}.       &28&Xeon+Virtex-7&12500& 2600             &  394\\
ESP32-WROOM-32 & 40 & Xtensa 32-bit LX6 & 2 & 14 & 7$\times$10$^4$\\
\hline
\end{tabular}
\end{table*}

\begin{table*}\centering
\caption{\label{t:hscd} Design Space Comparison}
\small
\begin{tabular}{@{}lccrrcc@{}}
\hline
\textbf{No.} & \textbf{\begin{tabular}[c]{@{}c@{}}Blocks\\ on HW\end{tabular}} & \textbf{\begin{tabular}[c]{@{}c@{}}Blocks \\ on SW\end{tabular}} & \multicolumn{1}{c}{\textbf{\begin{tabular}[c]{@{}c@{}}Runtime\\ {[}cycles{]}\end{tabular}}} & \textbf{\begin{tabular}[c]{@{}r@{}}Std Cell\\ Count\end{tabular}} & 
\textbf{\begin{tabular}[c]{@{}c@{}}Total Area\\ {[}mm$^2${]}\end{tabular}} & 
\multicolumn{1}{c}{\textbf{\begin{tabular}[c]{@{}c@{}}Power\\{[}mW{]}\end{tabular}}}\\ \hline
1 & - & Trellis + Traceback & 10511397 & 119884 & 0.355 & 19.0 \\
2 & Trellis & Traceback & 3332256 & 276450 & 0.750 & 18.4 \\
3 & Trellis + Traceback & - & 38532 & 282380 & 1.151 & 20.2 \\
\hline
\end{tabular}
\end{table*}

For the SoC's highest performance setting, a 200-MHz core+AccelB run, the energy efficiency is 1.1~Gevents/J or about 1~nJ per event processed, a value roughly equivalent to a DRAM fetch~\cite{Horowitz14}.  Accounting for the nearly three thousand arithmetic operations that need to be computed per event, the average accelerated overhead energy per operation is about 0.35~pJ.  For reference, commercially relevant narrowband-IoT (NB-IoT) devices consume between about 400-6000~$\mu$J per byte depending on wireless channel conditions~\cite{Michelinakis20IoT}.  Thus, the SoC's computational energy requirements are far below existing communications solutions.  This strengthens the case for providing more processing capability within DNA processors for IoT applications.  Our peak measured energy efficiency is around 3~Gevents/J, a value obtained at a clock of 87.5~MHz for the core+AccelB setting at which point these components could be powered from a 0.51-V supply.  

As with performance, the SoC in its core+AccelB setting possesses a significant advantage relative to its AccelA and core only counterparts on the energy efficiency metric.  For example, at 200-MHz, the advantage is 28$\times$ and 146$\times$, respectively.  Naturally, relative to the factor improvements seen for performance, as we progress from core only to AccelA, to AccelB, the overhead incurred by additional acceleration circuitry diminishes the relative ratios.  Nonetheless, the efficiency margins between these designs remain substantial.  The measured energy efficiencies of the commercial platforms included in Fig.~\ref{f:EE} show between about four and five order-of-magnitude factor disparities between the AccelB SoC and the Cortex-A53 and ESP-32 implementations (summaries are provided in Table~\ref{t:comp}) respectively.  In particular, at its highest frequency setting, a 1.2-GHz clock with a power overhead of 546~mW, the SIMD-enabled multicore Cortex A53 is 2934$\times$ less energy efficient than the SoC.

A tabulated comparison of the SoC to other hardware executing the detection algorithm is provided in Table~\ref{t:comp}.  Key metrics discussed above are shown including our SoC's peak 200-MHz event throughput of 2.6~Mevents/s which is nearly twice the throughput of the fastest of the platforms discussed above.  In Table~\ref{t:comp} we also note an FPGA-accelerated desktop system~\cite{Wu20} that can achieve detection speeds of about 4.8$\times$ greater than our chip, a capability facilitated by high clock speeds and a wide PCIe link between the desktop and its FPGA support.  Naturally, these additional needs imposed a greater power overhead relative to our system resulting in the nearly three orders of magnitude energy efficiency ratio listed in Table~\ref{t:comp}.

Table~\ref{t:hscd} evaluates the performance/cost trade-off of our SoC's alternative implementations from the perspective of the workload's partition across hardware and software. Allocating trellis construction to dedicated logic results in the aformentioned 3$\times$ speedup over a software only design but with a 2.3$\times$ increase in hardware resources, measured here as combinational and sequential standard cells. Thus, the penalty incurred in terms of logic is roughly two and a half RISC-V cores - a substantial cost for admittedly a slightly more substantial gain in detection speed. A nearly identical cost is incurred when performing traceback in hardware as well. However, the dramatic increase in detection speed more than justifies the other design costs. In fact, the all-custom-hardware solution only becomes less favourable when non-performance requirements cannot be met. In other words, only when area and/or design complexity constraints preclude an ASIC solution do hardware/software co-design considerations become critical. 

\section{Summary and Conclusions}\label{s:summary}
The physical footprints of DNA sequencing machines have been significantly reduced over the last ten years.  Critical components of these systems, in particular sensors and associated analog electronics, have been combined into portable, hand-sized devices.  As a result, the breadth of uses available to DNA sequencers (e.g., personalized medicine, environmental monitoring, etc.) is quickly expanding with many opportunities for application in IoT contexts.

Presently however, these miniaturized sequencers do not contain any substantial computing resources.  Rather, they are tethered to traditional computing platforms via relatively high-bandwidth connections.  This will continue to impede the exciting application potential of this technology as part of an IoT solution.  With customized computing resources however, the possibility exists to internalize critical calculations in the DNA signal processing pipeline.  This will not only make the sequencing devices more intelligent, but significantly reduce the amount of information they need to share with one another across IoT networks.

In this paper, we introduced a 22-nm FDSOI CMOS SoC as a model of the embedded computing potential available to emerging miniature DNA sequencers.  Our chip consists of a scalar RISC-V core with first-level split cache and two accelerators.  The accelerators combine to support a sequence detection algorithm, an early step in a potential DNA signal processing chain that predicts the text sequence associated with physical DNA measurements.  The accelerators communicate directly with memory via the core's cache and are programmed by the core via C-code using extended assembly instructions.  By employing heavily pipelined accelerator circuitry, aggressive dependence on loop unrolling, and an efficient memory/accelerator data exchange strategy our fastest SoC performs DNA sequence detection 13$\times$ faster than existing miniature sequencers can measure them.  This margin offers ample opportunity for the SoC to engage with other pertinent computations in possible future applications. Also critical is the energy efficiency with which our SoC can carry out its calculations.  The ability to run our system from a supplies as low as 0.5~V, employing dedicated calculations, using a low-complexity scalar core, and minimizing our data movement needs resulted in substantial power savings.  Even at our proposed SoC's peak performance setting while running from a 200-MHz clock, the overhead energy needed to complete all the calculations pertaining to a single event are about six orders of magnitude smaller than that needed to wirelessly communicate them. \\
\indent Thus, as currently designed, our SoC is meant to handle the first step of the sequencing pipeline and interface with downstream computational tasks like mapping, alignment, and variant calling through an edge-cloud computing model focused on efficient data transfer and format compatibility. The physical interface for data transfer would depend on the deployment scenario. For local transfer to a desktop, a high-speed wired connection such as a USB link could be used to transfer the generated data to a local workstation. For scenarios where immediate cloud processing is desired, a networked connection, such as Ethernet or Wi-Fi, would enable the ASIC-equipped system to directly upload data to cloud storage or data ingestion services. Further, to bridge the gap between the ASIC's output and cloud-based computational resources, a software layer would be essential. This would make the data readily accessible to the downstream mapping, alignment, and variant calling algorithms running on cloud-based compute clusters and high-performance virtual machines. As the ASIC outputs raw base calls in a structured format, its results can be readily converted to FASTQ. The communication protocols used to facilitate data exchange could include API calls for structured data transfer and control, message queues for reliable and asynchronous delivery of data chunks, or real time streaming protocols for continuous and low-latency data transfer, depending on the specific requirements.

In sum, by demonstrating a high-performance, low-power DNA sequence detection SoC, this work paves the way for developing portable, energy-efficient biomedical consumer electronics devices, including point-of-care diagnostic tools, wearable sequencing platforms, and other mobile health technologies.

\section*{Acknowledgment}
The authors would like to thank the SSHERC, NSERC, CMC Microsystems, and Qualcomm Canada Inc. for financial support.

\ifCLASSOPTIONcaptionsoff
  \newpage
\fi

\bibliographystyle{IEEEtran}
\bibliography{RefIOT23}

\end{document}